\documentclass[aps,pre,reprint]{revtex4-1} 

\usepackage{amsmath,amssymb}
\usepackage{bm}
\usepackage{color}
\usepackage{graphicx}
\usepackage{hyperref}
\usepackage[utf8]{inputenc}
\usepackage{xcolor}

\usepackage{Learning}

\begin{document}

\title{Thermodynamic efficiency of learning a rule in neural networks}
\author{Sebastian Goldt}
\email{goldt@theo2.physik.uni-stuttgart.de}
\author{Udo Seifert}
\affiliation{II.\ Institut für Theoretische Physik, Universität Stuttgart, 70550
  Stuttgart, Germany}
\date{\today}

\begin{abstract}
  Biological systems have to build models from their sensory data that allow
  them to efficiently process previously unseen inputs. Here, we study a neural
  network learning a linearly separable rule using examples provided by a
  teacher. We analyse the ability of the network to apply the rule to new
  inputs, that is to generalise from past experience. Using stochastic
  thermodynamics, we show that the thermodynamic costs of the learning process
  provide an upper bound on the amount of information that the network is able
  to learn from its teacher for both batch and online learning. This allows us
  to introduce a thermodynamic efficiency of learning. We analytically compute
  the dynamics and the efficiency of a noisy neural network performing online
  learning in the thermodynamic limit. In particular, we analyse three popular
  learning algorithms, namely Hebbian, Perceptron and AdaTron learning. Our work
  extends the methods of stochastic thermodynamics to a new type of learning
  problem and might form a suitable basis for investigating the thermodynamics
  of decision-making.
\end{abstract}

\maketitle

\section{Introduction}
\label{sec:introduction}

Biological information processing occurs in three steps. First, an organism
needs to acquire information about the external state of affairs by sensing its
environment, for example by monitoring the concentration of nutrients in the
surrounding solution. Second, a model or a representation of the data is built
to allow for efficient processing. Such a model would then be the basis for the
third and final step: processing previously unseen inputs by applying the model
and making decisions based on the model's output, \emph{i.e.} to ``generalise''
from past experience.

The first step, sensing, has a history of interest from physicists going back at
least to the seminal work of Berg and Purcell~\cite{berg1977} searching for the
fundamental limits on sensing imposed by physics. More recently, the application
of stochastic thermodynamics~\cite{seifert2012,parrondo2015}, an integrated
framework to analyse the interplay of dissipation and information processing in
fluctuating systems far from equilibrium, has provided some intriguing results
with regards to the physical limits on the speed, precision and dissipation of
sensing in living
systems~\cite{Qian2005,Keymer2006,Tu2008,endres2009,lan2012,mehta2012,Skoge2013,
  govern2014,govern2014a,barato2014a,lang2014,sartori2014,ito2015,tenwolde2016}.

Neural networks, well known from statistical physics and machine
learning~\cite{nishimori2001,engel2001,mackay2003}, form a mature framework to
investigate learning and generalising. We have recently introduced the methods
of stochastic thermodynamics to study the thermodynamic efficiency of the second
step, building an efficient representation of uncorrelated
data~\cite{goldt2017}, and a recent study has looked at the non-equilibrium
thermodynamics of unsupervised learning with restricted Boltzmann
machines~\cite{salazar2017}.

In this paper, we study the learning of a rule by a neural network. The rules we
want to learn are Boolean functions: they take an input, call it $\vxi$, and map
it to a binary output, the true label $\vxi\to\tlab=\pm1$. The network has to
build a model of this rule by looking at a number of pairs $(\vxi, \tlab)$. Our
focus is on the final step of information processing: how well can the network
emulate the function after a training period, \emph{i.e.} how well do the
outputs of the network, $\lab$, match the correct output of the function
$\tlab$? We will show that the ability of the network to generalise such a rule
from the examples it has seen to previously unseen inputs is bound by the
dissipation of free energy by the components of the network as a consequence of
the second law of stochastic thermodynamics.

Our results apply to a wide variety of learning algorithms. For illustration
purposes, we analyse three learning algorithms in particular: Hebbian
learning~\cite{hebb1949,vallet1989}, which was inspired by the neurobiology of
memory formation; the celebrated Perceptron algorithm~\cite{rosenblatt1962},
whose discovery led to a surge in interest in neural networks in the 1960s and
which is still very influential; and finally AdaTron learning~\cite{anlauf1989},
a refinement of the Perceptron algorithm with surprising dynamical features.

This paper is organised as follows. We give a detailed description of our model
and its dynamics in Sec.~\ref{sec:model} and~\ref{sec:dynamics}. We derive a
general bound in Sec.~\ref{sec:bound1} and discuss a number of simple examples
with different learning algorithms in Sec.~\ref{sec:examples1}. We then derive a
second, sharper bound in Sec.~\ref{sec:results2} and analyse the efficiency of
learning in large networks in Sec.~\ref{sec:examples2}. We give some concluding
perspectives in Sec.~\ref{sec:conclusion}. Detailed proofs and a number of
technical points are discussed in the appendices.

\section{Inputs and labels, Teacher and student}
\label{sec:model}

\begin{figure}
  \centering
  \includegraphics[width=\columnwidth]{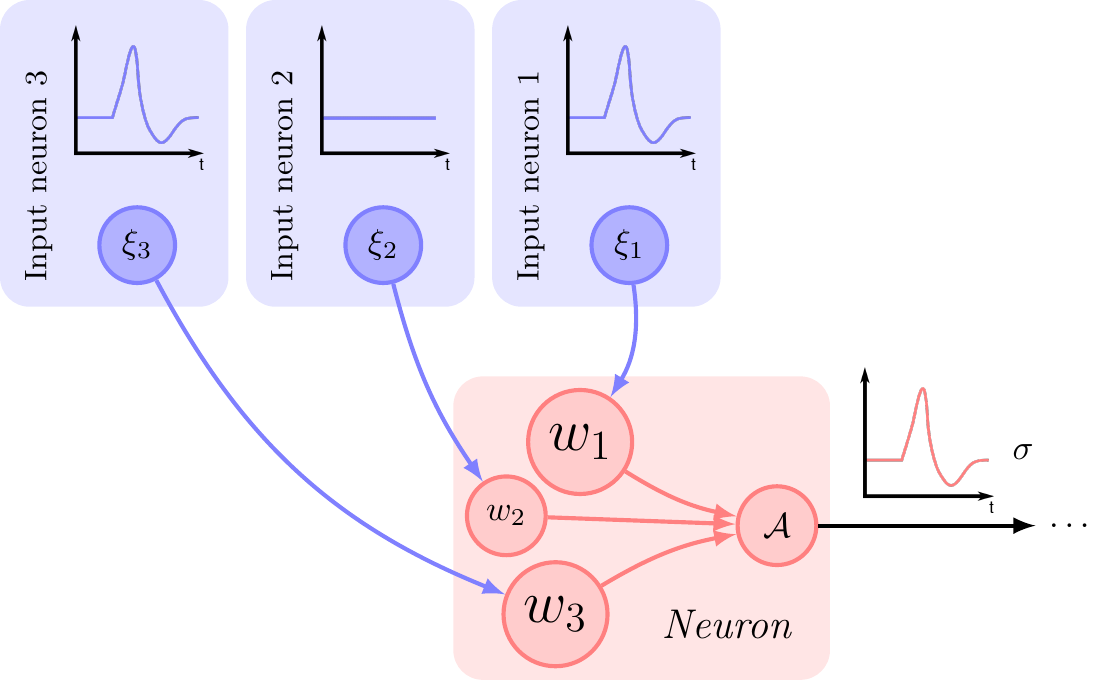}
  \caption{\label{fig:neural-network}\textbf{Snapshot in time of a simple neural
      network}. In this small network ($N=3$), the neuron of interest (red) has
    three input neurons (blue), two of which are just sending a short signal, an
    action potential, while the second input neuron is silent. This behaviour is
    captured by input vector~$\vxi=\set{1,-1,1}$ for the network shown. Each of
    the connections between the neuron of interest and an input neuron has a
    weight~$\w_n$. The neuron, which is fully characterised by its weights
    $\vw=(w_1,\dots,w_N)\in\mathbb{R}^N$, will also send an action potential,
    depending on its activation~$\act\sim\vw\cdot\vxi$. The response of the
    neuron is denoted by~$\lab$. In this example, $\lab=1$.}
\end{figure}

We consider a single neuron, modeled by a single-layer neural network shown
schematically in Fig.~\ref{fig:neural-network}. In neural networks, individual
neurons communicate via action potentials, \emph{i.e.} short electric pulses
that are the basic token of communication in biological neural
networks~\cite{kandel2000}. The neuron's task is to decide whether to fire an
action potential or not given an input it gets from the~$N$ neurons that it is
connected to. Since we are not interested in the precise temporal dynamics of
the action potentials, we model the input of the neuron, \emph{i.e.} the signals
that the neuron receives at a particular point in time, as
vectors~$\vxi=(\xi_1,\dots,\xi_N)$ where~$\xi_n=1$ if the~$n$th connected neuron
is firing an action potential in that
input. 
For symmetry reasons, we set~$\xi_n=-1$ if the~$n$th neuron is silent. The
inputs are distributed according to
\begin{equation}
  \label{eq:inputs}
  p(\vxi) = \prod_{n=1}^N\frac{1}{2}\left[ \delta(\xi_n-1)+\delta(\xi_n+1) \right].
\end{equation}

The neuron itself is fully characterised by the~$N$ weights~$\vw\in\mathbb{R}^N$
of its~$N$ afferent connections. The weights obey noisy dynamics, to be
specified in Sec.~\ref{sec:dynamics}. Presented with a given input~$\vxi$, the
neuron computes an input-dependent activation
\begin{equation}
  \label{eq:activation}
  \act \equiv \frac{1}{\sqrt{N}}\vw\cdot\vxi
\end{equation}
where the prefactor ensures normalisation. The activation determines whether the
neuron will fire an action potential or not, $\lab=1$ or $-1$, respectively. If
the prediction was noise-free, we would have $\lab=\sgn(\mathcal{A})$, where
$\sgn(x>0)=1$ and $\sgn(x\le0)=-1$; instead, the predicted label $\lab$ is
stochastic with
\begin{equation}
  \label{eq:lab}
  p(\lab|\mathcal{A}) \propto \exp(\beta \lab \act)
\end{equation}
where $\beta$ is the inverse temperature of the surrounding heat bath. We set
$k_B=\beta=1$ for the remainder of this article, rendering entropy and energy
dimensionless without loss of generality.

The rules we want to learn are Boolean functions of the inputs. More precisely,
we will focus on realisable rules which are linearly separable, \emph{i.e.} we
can write 
\begin{equation}
  \label{eq:tlab}
  \tlab = \sgn(\vT\cdot\vxi) = \pm 1
\end{equation}
where the teacher network $\vT\in\mathbb{R}^N$ has the same architecture as the
neural network $\vw$. The components of the teacher are independent and drawn
from a normal distribution with mean 0 and variance 1 and kept fixed. We draw
the teacher at random in order to make general statements about the ability of
the network to infer teachers of this form. By analogy, the neuron in such a
setup is often called the student. We can interpret the true label of an input
as an indication of whether the student should fire an action potential in
response to that input or not. We emphasise that while the response of a neuron
to an input is stochastic, as is the case physiologically, we assume that the
teacher does not make mistakes.

\begin{figure}
  \centering
  \includegraphics[width=.9\columnwidth]{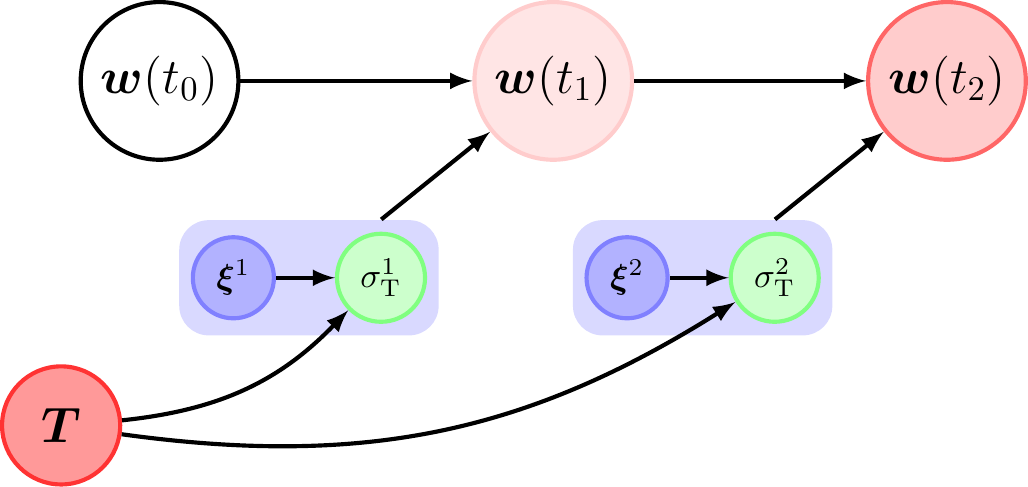}
  \caption{\label{fig:neuron}\textbf{A single neuron learning a rule.} The
    neuron, characterised by its weights $\vw\in\mathbb{R}^N$,
    is presented with a succession of inputs $\vxi^\mu\in\{-1,1\} ^N$
    and their true labels $\tlab[\mu]=\sgn(\vT\cdot\vxi^\mu)=\pm1$ which are
    determined by a random, static teacher $\vT\in\mathbb{R}^N$. The goal
    of learning is to infer the teacher $\vT$ by using only the information
    provided by the samples $(\vxi^\mu, \tlab[\mu])$, such that the neuron is
    eventually able to predict the true label of a previously unseen input. 
  }
\end{figure}

The goal of learning is to adjust the weights of the network $\vw$ such that the
label predicted by the neuron equals the true label for any input $\vxi$,
$\lab=\tlab$. The adaptation of weights is thought to be a main mechanism of
memory formation in biological networks~\cite{kandel2000}. To this end, the
neuron needs to infer the teacher $\vT$. However, the neuron only has indirect
access to the teacher via a number of samples $(\vxi^\mu, \tlab[\mu])$, where we
have now indexed the inputs and their labels with $\mu=1,\dots$, see
Fig.~\ref{fig:neuron}. The exact form of the dynamics will be specified below in
Section~\ref{sec:dynamics}. A classic example for neurons performing this kind
of associative learning are the Purkinje cells in the
cerebellum~\cite{Marr1969,Albus1971,ito1982a,ito1982b}.

\clearpage

\section{Dynamics}
\label{sec:dynamics}

Let us now describe the dynamics of the weights learning a rule from a fixed
teacher $\vT$. Initially, all the weights are independent of each other and in
equilibrium in the potential
\begin{equation}
  \label{eq:potential}
  V(\vw) = \frac{k}{2}\vw\cdot\vw
\end{equation}
which restricts the weights from increasing indefinitely, as is also the case
physiologically~\cite{dayan2001}.

Starting at time $t=0$, the weights $\w_n$ obey overdamped Langevin
equations~\cite{vankampen1992}
\begin{equation}
  \label{eq:langevin-short}
  \dot{\w}_n(t) = F_n\left(\vw(t), \tlab[\mu(t)], \vxi^{\mu(t)}, t\right) + \zeta_n(t).
\end{equation}
The thermal white noise $\zeta_n(t)$ has correlations
$\avg{\zeta_n(t)\zeta_m(t')}=2 D \delta_{nm}\delta(t-t')$, where $D$ is the
``diffusion'' constant. We set the mobility of the weights to unity and impose the
fluctuation-dissipation relation $\beta D = 1$ for thermodynamic
consistency~\cite{seifert2012}. For the remainder of this article, we will use
angled brackets $\avg{\cdot}$ to indicate averages over the thermal noise,
unless indicated otherwise.

The total force $\vec{F}=(F_1,\dots,F_N)$ on the weights has a conservative
contribution from the harmonic potential, $-\nabla V(\vw)=-k\vw$, and a
non-conservative contribution from the learning force $\vec{f}$, which is a
function of a single sample $(\tlab[\mu(t)], \vxi^{\mu(t)})$. The learning force
changes the weights in such a way that the neuron becomes more likely to predict
the true label for the input as discussed above. In this paper, we focus on
\emph{online learning}, where the learning force changes the weights using just
a single sample at a time. The succession of samples is described by the
function $\mu(t)=1,2,\dots$. This function may be deterministic or stochastic
and we do not make any assumptions about the rate of change of the inputs nor
whether the same input may be shown more than once to the neuron. We note that
our results also hold for \emph{batch learning}, where the neuron has
simultaneous access to a set of samples at any point in time as discussed in
detail in Appendix~\ref{sec:batch-learning}.

We will assume that the change to the weights in response to a sample is made in
the direction of that input, as is the case for most customary algorithms (see
Sec.~\ref{sec:examples1} and~\cite{mace1998,engel2001,mackay2003}). 
We thus write $\vec{f}=(f_1,\dots,f_N)$ with
\begin{equation}
  \label{eq:learning-force}
  f_n \equiv \lr(t) \xi_n^{\mu(t)} \tlab^{\mu(t)} \mathcal{F}(|\vw(t)|, \vw(t)\cdot\vxi^{\mu(t)}, \vT\cdot\vxi^{\mu(t)}),
\end{equation}
where we have introduced a possibly time-dependent learning rate $\lr(t)$
\footnote{We emphasise that the learning rate that we denote $\lr(t)$ in this
  paper is an established concept in the analysis of neural networks and should
  not be confused with the learning rate
  $l_n$~\cite{allahverdyan2009,hartich2014} or the information
  flow~\cite{horowitz2014,horowitz2015} from stochastic thermodynamics which we
  use to prove our main result (see Appendix~\ref{sec:derivation1}). Here, we
  will use the term ``thermodynamic learning rate'' to refer to the latter.} and
we denote the Euclidean norm of a vector by $|\cdot|$. Here, $\mathcal{F}$ is an
as yet unspecified scalar function of the length of the weight vector,
$|\vw(t)|$, the student's field $\vw(t)\cdot\vxi^{\mu(t)}$ and the teacher's
field $\vT\cdot\vxi^{\mu(t)}$. The learning force may only depend on the sign of
the teacher's field. Its precise form is specified by learning algorithms; some
popular forms are summarised in Tab.~\ref{tab:rules} and described in more
detail in Sec.~\ref{sec:examples1}. However, we stress that the bounds that we
derive in this paper do not depend on the particular form of the learning force
and hold for all learning dynamics of the form~\eqref{eq:langevin-short}.  The
full Langevin equation for a weight then reads
\begin{widetext}
  \begin{equation}
    \label{eq:langevin}
    \dot{\w}_n(t) = -k \w_n(t)\\
    + \lr(t) \xi^{\mu(t)}_n \tlab^{\mu(t)} \mathcal{F}(|\vw(t)|,
    \vw(t)\cdot\vxi^{\mu(t)}, \vT\cdot\vxi^{\mu(t)}) + \zeta_n(t).
  \end{equation}

On the ensemble level, the system is fully described by the distribution
$p(\vT, \vw, t)$. Its equation of motion is given by a Fokker-Planck
equation~\cite{vankampen1992} whose form is simplified by the fact that the
noise $\zeta_n(t)$ of the different weights is uncorrelated. The dynamics are
hence multipartite~\cite{hartich2014,horowitz2015} and the Fokker-Planck
equation corresponding to the Langevin dynamics~\eqref{eq:langevin} separates
into one probability current for every weight $\w_n$,
\begin{equation}
  \label{eq:fpe}
  \partial_t p(\vT, \vw, t)=-\sum_n^N \partial_n j_n(\vT, \vw, t),
\end{equation}
where $\partial_t\equiv \partial/\partial t$,
$\partial_n\equiv \partial/\partial \w_n$ and the probability currents are given
by
\begin{equation}
  \label{eq:jn}
  j_n(\vT, \vw, t) = \left[-k \w_n + \lr(t) \xi_n^{\mu(t)} \tlab^{\mu(t)}
    \mathcal{F}(|\vw|, \vw\cdot\vxi^{\mu(t)}, \vT\cdot\vxi^{\mu(t)})\right]p(\vT,
  \vw, t) - D\partial_n p(\vT, \vw, t).
\end{equation}
\end{widetext}
There are hence three sources of stochasticity in the system. On the one hand,
the fluctuating weights $\w(t)$ and the stochastic process of firing an action
potential or not, $\lab$, for a given activation~\eqref{eq:activation} affect
the performance of the network. Furthermore, there is randomness in the choice
of samples during learning. Since the neuron learns using just a single randomly
drawn input and its label at a time, the system performs stochastic gradient
descent in the sense that the direction of the learning force fluctuates from
one input to the next and only yields the appropriate direction for the weights
on average.

\clearpage

\section{A first thermodynamic bound on generalising}
\label{sec:bound1}

The aim of the neuron is to predict the label of a previously unseen input
$\vxi$ as well as possible. In the following discussion, we consider the
generalisation properties of the neuron, \emph{i.e.} its performance on an input
drawn at random from the distribution~\eqref{eq:inputs}, so we drop the
superscript $\mu$ on inputs and labels. We quantify the accuracy of the
predictions using information theory~\cite{mackay2003,cover2006a}. The \emph{a
  priori} uncertainty about the true label of some input, $\tlab$, is quantified
by the Shannon entropy of the random variable $\tlab$ and defined as
\begin{equation}
  \label{eq:shannon}
  S(\tlab) \equiv -\sum_{\tlab=\pm1} p(\tlab)\ln p(\tlab)=\ln 2,
\end{equation} 
where the equality follows from the fact that for an arbitrary input with
probability distribution~\eqref{eq:inputs}, $p(\tlab=\pm1)=1/2$. We have
included the variable $\tlab$ as an argument to the Shannon entropy in a slight
abuse of notation to emphasise that $S(\tlab)$ is the entropy of the
\emph{marginalised} distribution of just $\tlab$. The
definition~\eqref{eq:shannon} readily carries over to continuous random
variables, where the sum is replaced by an integral over the support of the
random variable.

The uncertainty about the true label given the label predicted by the neuron
$\lab$ is given by the conditional Shannon entropy
\begin{align}
  \label{eq:shannon-conditional}
  S(\tlab|\lab)\equiv&-\sum_{\tlab,\lab}p(\tlab,\lab)\ln p(\tlab|\lab) \\
  & \le S(\tlab)
\end{align}
where $p(\tlab|\lab)=p(\tlab,\lab)/p(\lab)$ and the inequality indicates that,
on average, knowing the label predicted by the neuron reduces uncertainty about
its true label.

The natural quantity to measure the information learnt by the neuron is the
mutual information
\begin{equation}
  \label{eq:mutual-info}
  \mutual{\tlab}{\lab} \equiv S(\tlab) - S(\tlab|\lab) \ge 0
\end{equation}
which measures by how much, on average, the uncertainty about $\tlab$ is reduced
by knowing $\lab$. If learning and predicting went perfectly, then by knowing
the neuron's output $\lab$ one could predict the true label $\tlab$ with perfect
accuracy, such that $S(\tlab|\lab)=0$ and hence $\mutual{\tlab}{\lab}=\ln 2$. On
the other hand, when the weights are in equilibrium in their potential $V(\vw)$
before learning, there is no correlation between the weights of the student and
those of its teacher, such that $\mutual{\tlab}{\lab}=0$.

We can connect the mutual information $\mutual{\tlab}{\lab}$ to the
well-established generalisation error $\epsilon$ of neural
networks~\cite{mace1998,engel2001}. It gives the probability that the
neuron predicts the wrong label for an arbitrary input $\vxi$, assuming that the
prediction of the neuron is noise-free, \emph{i.e.}~$\lab=\sgn(\vw\cdot\vxi)$,
and is defined as
\begin{equation}
  \label{eq:epsilon}
  \epsilon = \avg{\theta(-\vw\cdot\vT)}
\end{equation}
where $\theta$ is the Heaviside step function. If the neuron predicted a label
based on its activity reliably via $\lab=\sgn(\vw\cdot\vxi)$ like the teacher,
Eq.~\eqref{eq:tlab}, the mutual information between the true and predicted label
for an arbitrarily drawn input could be expressed as
\begin{equation}
  \label{eq:mutual-info-labels}
  \mutual{\tlab}{\lab} = \ln 2 - S(\epsilon),
\end{equation}
where $S(p)=-p \ln p-(1-p)\ln (1-p)$ is the shorthand for the Shannon entropy of
a binary stochastic variable with probability $p$. For a realistic neuron, the
activity gives only the probability that the neuron will fire an action
potential, see Eq.~\eqref{eq:lab}, hence Eq.~\eqref{eq:mutual-info-labels}
constitutes an upper bound on its actual performance with noisy predictions. In
the following, we will focus on deriving thermodynamic bounds on the amount of
information that the neuron can learn from its teacher for the ideal case of
noise-free predictions.

Thermodynamics enters the picture by considering the free energy costs of the
non-equilibrium dynamics of the weights. They can be quantified by the total
entropy production $\Delta S^\tot_n$ of a single weight in the network which is
guaranteed to be non-negative by the second law of stochastic thermodynamics and
has two contributions: the heat dissipated by the $n$th weight into the
connected heat bath, $\Delta Q_n$, and the change in Shannon entropy of the
marginalised distribution $p(\w_n)$~\cite{seifert2012}.

For a neural network learning with the dynamics~\eqref{eq:langevin}, we can show
both for $N=1$ and in the thermodynamic limit that
\begin{equation}
  \label{eq:inequality1}
  \mutual{\tlab}{\lab} \le \Delta S^\tot_n \equiv \Delta S(\w_n) + \Delta Q_n
\end{equation}
from the second law for the network (see Appendices~\ref{sec:derivation1}
and~\ref{sec:max-ent} for details). This suggests the introduction of an
efficiency
\begin{equation}
  \label{eq:efficiency}
  \eta\equiv \frac{\mutual{\tlab}{\lab}}{\Delta S(\w_n) + \Delta Q_n} \le 1.
\end{equation}
This inequality is our first main result and holds at all times $t>0$ in
Eq.~\eqref{eq:langevin} and~\eqref{eq:fpe}.

We note that while this result is superficially similar to the inequality we
have derived previously~\cite{goldt2017}, here we consider an entirely different
scenario. In our previous work, there was no teacher; instead, we considered the
learning of a number of \emph{fixed} inputs with true labels drawn at random,
such that the true labels were uncorrelated to the inputs and to each
other. Hence the concept of a generalisation error did not apply and ``the
information'' was always related to the labels of the fixed set of inputs. Here,
we are learning from a number of samples $(\tlab[\mu], \vxi^\mu)$ which are
examples of the function that the teacher performs, Eq.~\eqref{eq:tlab}. The
network tries to infer this function in order to be able to correctly classify
\emph{previously unseen} inputs. What we show here is that the ability to learn
from a teacher and generalise accordingly is bound by the total entropy
production per weight.

\begin{table}
  \begingroup 
  \def\arraystretch{1.4}
  \begin{tabular}{lcc}
    \emph{Algorithm}           & $\mathcal{F}(|\vw|, \vw\cdot\vxi^\mu, \tlab[\mu])$ &  \emph{Ref. } \\
    \toprule
    \square{hebbian} Hebbian    & 1 &\cite{hebb1949,vallet1989}\\
    \square{perceptron} Perceptron & $\theta(-\tlab[\mu] \vw\cdot\vxi^\mu)$ &\cite{rosenblatt1962,biehl1994} \\
    \square{adatron} AdaTron  & $\vw\cdot\vxi^\mu/\sqrt{N}\,\theta(-\tlab[\mu]\vw\cdot\vxi^\mu)$
                                                                                    &\cite{anlauf1989} \\
  \end{tabular}
  \caption{\label{tab:rules}\textbf{Different learning algorithms} for a neuron
    with weights $\vw$ online-learning a sample $(\tlab[\mu], \vxi^\mu)$
    together with the colour code used throughout the paper. Here,
    $\theta(\cdot)$ is the Heaviside step function. References are given to
    where the algorithm first appeared in a discussion of (the statistical
    mechanics of) neural learning, to the best of our knowledge. A detailed
    discussion of the form of these algorithms is given in
    Sec.~\ref{sec:examples1}}
  \endgroup
\end{table}

\section{Efficiency of different learning algorithms with $N=1$}
\label{sec:examples1}

Let us look at a toy model of a neuron with a single weight. The weight is
initially in equilibrium in the harmonic potential $V(w)=kw^2/2$. Without loss
of generality, we can set $k=\beta=D=1$, making energy, entropy, time and the
weights dimensionless. At time $t=0$, the learning rate is suddenly increased
from $0$ to a constant value $\lr_0$.

The neuron learns using one of the three learning algorithms, each defined by a
particular choice of $\mathcal{F}$ and summarised in Tab.~\ref{tab:rules}. The
simplest non-trivial choice is $\mathcal{F}=1$, which is Hebbian
learning~\cite{hebb1949,vallet1989}. For such an algorithm, each incoming sample
changes the weight by an amount $\sim\tlab[\mu]\xi^\mu$. An obvious improvement
on this simple algorithm is to only change the weight if the network would
currently predict the wrong label for that input, which is achieved by choosing
$\mathcal{F}=\theta(-\tlab[\mu]\w\xi^\mu)$. This is the Perceptron
algorithm~\cite{rosenblatt1962}. A further refinement of this rule is achieved
by choosing $\mathcal{F}=|\w\xi^\mu|\theta(-\tlab[\mu]\w\xi^\mu)$ such that the
change in the weights is proportional to the confidence of the neuron in its
decision, measured by $|\w\xi^\mu|$.

The key insight to solve the dynamics in each case is that in one dimension,
$\tlab[\mu] \xi^\mu= \sgn(T)$ for all $\xi^\mu$, which is readily verified. This
has the appealing consequence that it is possible to rewrite the Langevin
equations for all three learning rules without any mention of the inputs
$\xi^\mu$. Instead, learning a rule is equivalent to a quench of the potential
of the weight from the simple harmonic form $V(\w)=\w^2/2$ to a new
$T$-dependent potential $V^\q(T, w)$, the exact form of which depends on the
learning algorithm chosen. They read
\begin{equation}
  \label{eq:potentials-n1}
  V^\q(T, w) = 
  \begin{cases}
    w^2/2 - \lr_0 \w\sgn(T) \\
     w^2/2 - \lr_0 \w \sgn(T)\theta(-\w T)\\
    w^2/2 (1-\lr_0 \,\sgn(T)\sgn(w) \theta (-w \,\sgn(T)))
  \end{cases}
\end{equation}
for Hebbian, Perceptron and AdaTron learning, respectively. The weight then
relaxes to the new equilibrium distribution, which is given by the Boltzmann
distribution. The heat dissipated by the weight during this isothermal
relaxation is given by
\begin{equation}
  \label{eq:deltaQ_N1}
  \Delta Q=\avg{V^\q(T, w)}_0-\avg{V^\q(T, w)}_\eq
\end{equation}
where $\avg{\cdot}_0$ and $\avg{\cdot}_\eq$ indicate averages with respect to
the distributions of teacher and weight at $t=0$ and after relaxation,
respectively.

\begin{figure}
  \centering
  \includegraphics[width=\columnwidth]{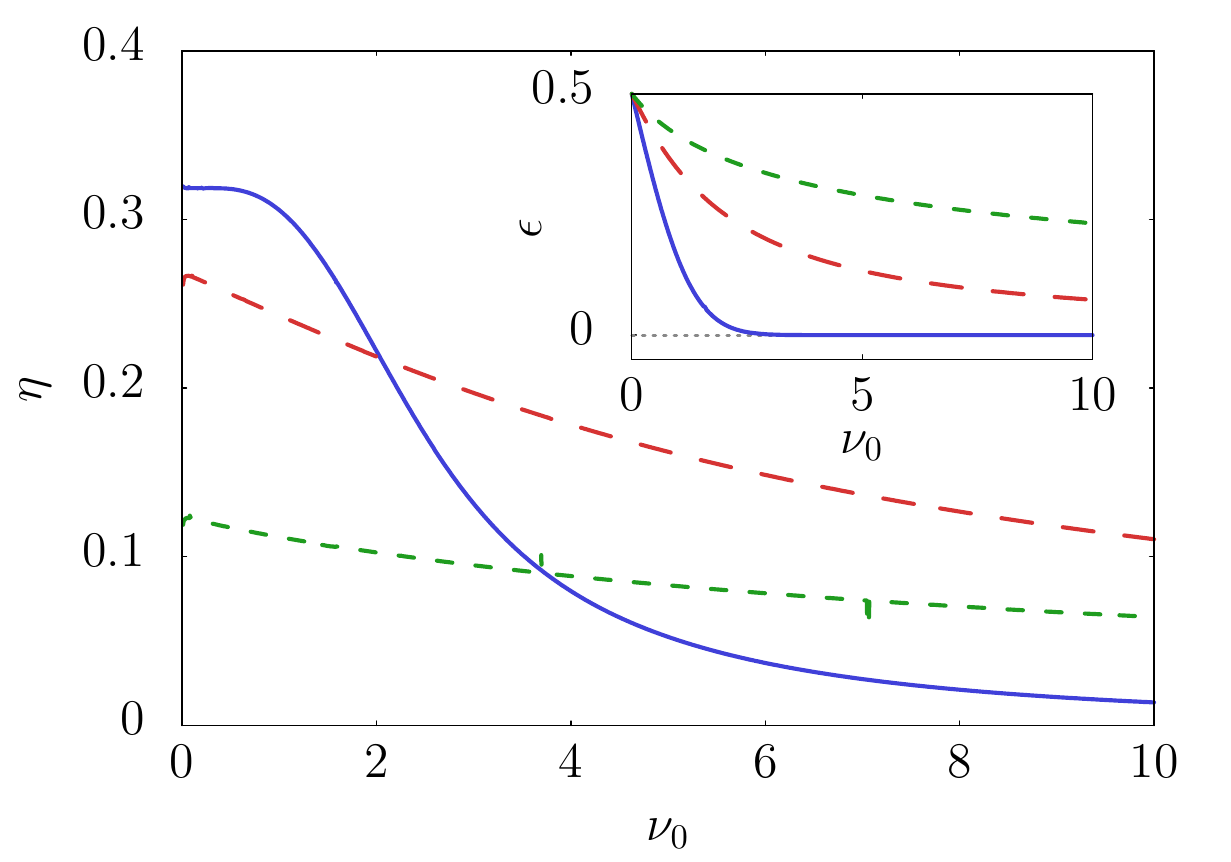}
  \caption{\label{fig:efficiency-n1}\textbf{Efficiency of a toy model with
      $N=1$.} We plot the efficiency $\eta$, Eq.~\eqref{eq:efficiency} and in
    the inset the generalisation error $\epsilon$, Eq.~\eqref{eq:epsilon}, as a
    function of the fixed learning rate $\lr_0$ for a neuron with a single
    weight, $N=1$, learning using the Hebbian~\square{hebbian} (solid),
    Perceptron~\square{perceptron} (long dashed) and AdaTron~\square{adatron}
    (dashed) algorithms. Parameters: $k=\beta=D=1$ without loss of generality.}
\end{figure}

We plot the efficiency of learning~\eqref{eq:inequality1} for $t\to\infty$ in
Fig.~\ref{fig:efficiency-n1} as a function of the learning rate. While the
Hebbian algorithm yields the lowest generalisation error, its efficiency is
quickly dominated by the heat dissipated, $\Delta Q\sim\lr_0^2$, resulting in
low efficiency. The perceptron algorithm is the most efficient for large $\lr$
and yields a better generalisation performance than the AdaTron algorithm, too
(see the inset of Fig.~\ref{fig:efficiency-n1}).

We finally note that our inequality~\eqref{eq:inequality1} is sharp for
$N=1$. Optimal efficiency $\eta\to1$ can for example be reached for Hebbian
learning with a time-dependent learning rate $\lr(t)$ where we first linearly
increase $\lr(t)$ from $0$ to $\lr_0$ over a period of time $\tau$ and then keep
it at the final value $\lr_0$:
\begin{equation}
  \label{eq:lr-time-dependent}
  \lr(t) \equiv
  \begin{cases}
    \lr_0 t / \tau & t < \tau \\
    \lr_0 & t \ge \tau,
  \end{cases}
\end{equation}
which is similar to an example discussed in~\cite{goldt2017}. In the limit of
slow driving $\tau\to\infty$, the dissipated heat $\Delta Q\to 0$. If
additionally the learning rate $\lr\to\infty$, the efficiency $\eta\to1$.

\section{Learning in large networks and a second bound}
\label{sec:results2}

We just saw in Sec.~\ref{sec:examples1} that for $N=1$,
$\xi^\mu\sgn(T \xi^\mu)=\sgn(T)$ which simplifies the analysis because the
inputs $\xi^\mu$ do not appear explicitly in the equation of motion of the
weight. In higher dimensions, we have instead a learning force on the $n$th
weight
\begin{equation}
  \label{eq:tlab_noisy}
  f_n \sim \xi_n^\mu\sgn(\vT\cdot\vxi^\mu) = \xi_n^\mu\sgn\left(T_n \xi^\mu_n+ \sum_{m\neq n}T_m \xi_m^\mu\right)
\end{equation}
which will fluctuate between the desired value $\sgn(T_n)$ and $-\sgn(T_n)$ due
to the second term inside the sign function, which is effectively a noise term
corrupting the signal from the $n$-th component of the teacher. So instead of
relaxing to a new equilibrium as seen for $N=1$, the weights relax to a steady
state with a constant, positive rate of thermodynamic entropy
production~\cite{seifert2012}. Our inequality~\eqref{eq:inequality1} still
applies to this process, but it is not very sharp anymore:
$\mutual{\tlab}{\lab}\sim1$ and $\Delta S(\w_n)\sim1$, but a steady state comes
with a non-zero rate of heat dissipation, such that $\Delta Q \sim t$. This
issue was not addressed in our previous work~\cite{goldt2017}. In this section,
we derive a sharper bound using concepts from steady state
thermodynamics~\cite{oono1998}.

We start with the explicit expression for the total entropy production of the
weights of the network~\cite{seifert2012},
\begin{equation}
  \label{eq:Stot-explicitly}
  \dot{S}^\tot(t) = \sum_n \int \dd \vT \dd \vw \; \frac{p(\vT, \vw, t)}{D}
  \left( \frac{j_n(\vT, \vw, t)}{p(\vT, \vw, t)} \right)^2 \ge 0.
\end{equation}
In our problem, the learning rate $\lr(t)$ acts as a control parameter. For
every value of $\lr(t)$, there is a well-defined steady state
$p^\st(\vT, \vw; \lr(t))$ where $\partial_t p^\st(\vT, \vw; \lr(t))=0$ as in
equilibrium, but where at least some of currents
$j_n^\st(\vT, \vw; \lr(t))\neq 0$, leading, \emph{inter alia}, to a constant
rate of total entropy production $\dot{S}^\tot\ge0$ in the steady state. For the
remainder of this section, we will use the shorthands
\begin{align}
    p =      & p(\vT, \vw, t),       &   p^\st=   & p^\st\left(\vT, \vw; \lr(t)\right), \\
    j_n=     & j_n(\vT, \vw, t), & j_n^\st= & j_n^\st\left(\vT, \vw; \lr(t)\right),
\end{align}
to keep our notation slim. We can rewrite the total entropy
production~\eqref{eq:Stot-explicitly},
\begin{widetext}
  \begin{align}
    \dot{S}^\tot(t) =&  \sum_n \int \dd \vT \dd \vw \; \frac{p}{D}
                       \left( \frac{j_n}{p} - \frac{j_n^\st}{p^\st} +
                       \frac{j_n^\st}{p^\st}\right)^2 & & \\
    = & \sum_n \int \dd \vT \dd \vw \; \frac{p}{D}\left( \frac{j_n}{p}
        -\frac{j_n^\st}{p^\st}\right)^2  + \sum_n \int \dd \vT \dd \vw \; \frac{p}{D} \left(
        \frac{j_n^\st}{p^\st} \right)^2  + 2\sum_n \int \dd \vT \dd \vw \;
        \frac{p}{D}\frac{j_n^\st}{p^\st}\left(
        \frac{j_n}{p}-\frac{j_n^\st}{p^\st} \right).
  \end{align}
\end{widetext}
The cross-term is identically zero, which can be seen by first rewriting the
integrand and then integrating by parts,
\begin{align}
  & \sum_n \int \dd \vT \dd \vw \;
    \frac{p}{D}\frac{j_n^\st}{p^\st}\left(\frac{j_n}{p}-\frac{j_n^\st}{p^\st}
    \right)\\
  =& \sum_n \int \dd \vT \dd \vw \; \frac{p}{D}\frac{j_n^\st}{p^\st} D
     (-\partial_n \ln \frac{p}{p^\st})\\
 =& \sum_n \int \dd \vT \dd \vw \;\partial_n j_n^\st \frac{p}{p^\st}  = 0,
\end{align}
where we have used the fact that our system is infinite to integrate and the
Fokker-Planck equation in the last step. We can hence
write~\cite{seifert2012,esposito2010b}
\begin{equation}
  \dot{S}^\tot(t) = \dot{S}^\na(t) + \dot{S}^\ad(t)
\end{equation}
where we have introduced the non-adiabatic entropy production
\begin{equation}
  \dot{S}^\na(t) \equiv \sum_n \int \dd \vT \dd \vw \; \frac{p}{D}\left( \frac{j_n}{p}
    -\frac{j_n^\st}{p^\st}\right)^2 \ge 0
\end{equation}
and the adiabatic entropy production
\begin{equation}
  \dot{S}^\ad(t) \equiv\sum_n \int \dd \vT \dd \vw \; \frac{p}{D} \left(
    \frac{j_n^\st}{p^\st} \right)^2\ge 0
\end{equation}
Both entropy production rates are evidently positive. They each correspond to a
possible mechanism that leads to the breaking of time symmetry and hence to
dissipation: the application of non-equilibrium constraints ($\dot{S}^\ad$) and
the presence of driving ($\dot{S}^\na$).

The non-adiabatic entropy production of the system can be written
as~\cite{esposito2010b}
\begin{equation}
  \label{eq:Sna}
  \dot{S}^\na(t) = - \int \dd \vT \dd \vw \;
  \dot{p}(\vT, \vw, t) \ln \frac{p(\vT, \vw, t)}{p^\st(\vT, \vw, \lr(t))}.
\end{equation}
It becomes identically zero once the steady state is reached, as is easily seen
from its definition. By splitting the logarithm, we find the second law of
steady state thermodynamics~\cite{oono1998,hatano2001,esposito2010b}
\begin{equation}
  \label{eq:2nd-law-sst}
  \dot{S}^\na(t) = \dot{S}(\vT, \vw, t) + \dot{Q}^\ex(t) \ge 0
\end{equation}
where $\dot{S}(\vT, \vw, t)$ is the rate of change of the Shannon entropy of the
distribution $p(\vT, \vw, t)$ and we have identified the excess
heat~\cite{oono1998,esposito2010b,hatano2001}
\begin{equation}
  \label{eq:Sex}
  \dot{Q}^\ex(t) \equiv \int \dd \vT \dd \vw \; \dot{p}(\vT, \vw, t) \;
  \ln p^\st(\vT, \vw, \lr(t))
\end{equation}
which has no definite sign.

Starting from the second law of steady-state
thermodynamics~\eqref{eq:2nd-law-sst}, we can derive our second, sharper bound
on the accuracy of learning:
\begin{equation}
  \label{eq:inequality2}
  \mutual{\tlab}{\lab} \le \Delta S(\w_n) + \Delta Q^\ex_n 
\end{equation}
which leads to the efficiency
\begin{equation}
  \label{eq:efficiency2}
  \tilde{\eta} \equiv \frac{\mutual{\tlab}{\lab}}{\Delta S(\w_n) + \Delta
    Q^\ex_n} \le 1.
\end{equation}
This the second main result of our paper. It also holds at all times and applies
to any learning algorithm that depends on the weights, $\vw$, and samples
$(\tlab^{\mu(t)}, \vxi^{\mu(t)})$. We give the details of its derivation in
Appendix~\ref{sec:derivation2} and show that our result applies to batch
learning in Appendix~\ref{sec:batch-learning}.

\section{Online learning in large networks}
\label{sec:examples2}

The number of afferent connections to a single neuron in a realistic network may
be on the order of thousands~\cite{kandel2000}, so it is sensible to analyse
learning in the limit $N\to\infty$. We will focus on online
learning~\cite{heskes1991,biehl1994} using the algorithms introduced in
Sec.~\ref{sec:examples1} and summarised in Table~\ref{tab:rules}. We will assume
that the samples, indexed by $\mu(t)$, change much faster than the weights
relax. This assumption is central to virtually all of the existing literature on
the analysis of online learning algorithms.

\subsection{Scaling of the learning rate}
\label{sec:scal-learn-rate}

We have noted that for $N>2$, the learning force on the $n$th weight will
fluctuate between two values proportional to $\pm\sgn(T_n)$, leading to a steady
state with constant $\epsilon$ and constant rate of heat dissipation. Let us try
to make this statement more quantitative by looking at the learning force
averaged over the inputs $\vxi$ in the limit $N\to\infty$. Setting
$\mathcal{F}=1$ for the moment for simplicity of notation, we have
\begin{equation}
  f_n = \lr(t) \tlab[\mu] \xi^\mu_n  = \lr(t) \xi^\mu_n \sgn(T_n\xi^\mu_n + \psi)
\end{equation}
where we have written $\mu=\mu(t)$ to simplify our notation and we have
introduced the noise term inside the $\sgn(\cdot)$ function,
\begin{equation}
  \psi\equiv\sum_{m\neq n}T_m \xi_m^\mu.
\end{equation}
$\psi$ is uncorrelated with $T_n\xi^\mu_n$ and normally distributed with zero mean
and variance $N-1\approx N$ due to the central limit theorem since the teacher
and the inputs are uncorrelated. We are interested in the probability
$p_\parallel$ that
\begin{equation}
  \label{eq:1}
  \sgn(T_n \xi_n^\mu + \psi) = \sgn(T_n\xi_n^\mu),
\end{equation}
\emph{i.e.} the probability that the learning force points in the right
direction despite the noise term $\psi$. This probability is found by
integrating the binormal distribution $p(T_n,\psi)=p(T_n)p(\psi)$ over the
region where~\eqref{eq:1} holds for $\xi_n^\mu=1$ and $\xi_n^\mu=-1$,
respectively. We find that
\begin{equation}
  \avg{f_n}_{\vxi} = \lr(t) (2 p_\parallel-1)\sgn(T_n) \sim \lr(t)
  \frac{\sgn(T_n)}{\sqrt{N}}
\end{equation}
where we have expanded $p_\parallel$ for large $N$~\cite{nhmf2010}. Hence the
larger the network, the smaller the information that $\tlab=\sgn(\vT\cdot\vxi)$
carries about a single component of the teacher network. This analysis suggests
we choose a learning rate $\lr(t)\equiv\lrn_0 \sqrt{N}$ with the normalised
learning rate $\lrn(t)\sim1$. This choice corresponds to the conventional
scaling of time with the inverse of the network size in the machine learning
literature~\cite{mace1998,engel2001}, which amounts to nothing more but an
increase in samples shown to the network to compensate for the dilution of the
signal.

\subsection{Dynamics}
\label{sec:examples2-dynamics}

First of all, we would like to compute the time-dependent generalisation error
$\epsilon(t)$ for online learning with the three algorithms from
Tab.~\ref{tab:rules} in a large network with dynamics given
by~\eqref{eq:langevin}. We keep the inverse temperature and the diffusion
constant at $\beta=D=1$ and again consider the case where the learning rate is
quenched to a constant value $\lrn=\lrn_0$ at $t=0$, leaving us with two free
parameters: $\lrn_0$ and the stiffness of the harmonic potential $k$, see
Eq.~\eqref{eq:potential}.

We thus introduce two new parameters, which go back to the original proof of
convergence of the perceptron algorithm~\cite{minsky1969} and play an important
role in the statistical mechanics of learning~\cite{engel2001},
\begin{equation}
  \label{eq:Q-and-R}
  \mathcal{Q} \equiv \frac{\vw\cdot\vw}{N} \quad \text{and} \quad  \mathcal{R} \equiv \frac{\vT\cdot\vw}{N}.
\end{equation}
These quantities have the appealing property of being self-averaging in the
thermodynamic limit, where they become the second moment of $\w_n$ and the
covariance of $(T_n, \w_n)$, respectively. Using geometrical~\cite{engel2001} or
analytical~\cite{mace1998} arguments, it can be shown that the generalisation
error~\eqref{eq:epsilon} becomes
\begin{equation}
  \label{eq:epsilon-thermo}
  \epsilon = \frac{1}{\pi}\arccos\left(\frac{\vw\cdot\vT}{|\vw||\vT|}\right) = \frac{1}{\pi}\arccos\left( \frac{\mathcal{R}}{\sqrt{\mathcal{Q}}} \right).
\end{equation}
Hence it is sufficient to find and solve the equations of motion for
$\mathcal{Q}$ and $\mathcal{R}$ to solve the dynamics of the generalisation
error.

We can indeed derive such equations directly from the Langevin equation for the
weights $\vw$~\eqref{eq:langevin} (see Appendix~\ref{sec:dynamics-sde}). They
read
\begin{subequations}
  \label{eq:eom}
  \begin{align}
    \begin{split}
      \dot{\mathcal{Q}} =& 2(1-k \mathcal{Q})  + 2 \lrn_0 \avg{\sgn(x)y\mathcal{F}(x, y)}_{\vxi} \\
                         & \qquad + \lrn_0^2 \avg{\mathcal{F}^2(x, y)}_{\vxi},
    \end{split}\\
    \dot{\mathcal{R}} =& -k \mathcal{R}  + \lrn_0\avg{\sgn(x)x\mathcal{F}(x, y)}_{\vxi},
  \end{align}
\end{subequations}
where we have introduced the auxiliary random variables
\begin{equation}
  \label{eq:x-and-y}
  x\equiv \vT\cdot\vxi/\sqrt{N}\qquad \text{and} \qquad y\equiv\vw\cdot\vxi/\sqrt{N}.
\end{equation}
Since we are assuming that the inputs change on a timescale much faster than the
relaxation time of the weights, we need to average Eqs.~\eqref{eq:eom} over the
inputs $\vxi$. This average is simplified by noting that the inputs only enter
the equations via $x$ and $y$. Thus the average over the inputs can be replaced
with an average over $x$ and $y$, which are binormally distributed by the
central limit theorem, with moments
\begin{equation}
  \begin{gathered}
    \avg{x}_\vxi=\avg{y}_\vxi=0,\\
    \avg{x^2}_\vxi=1, \quad \avg{y^2}_\vxi=\mathcal{Q}, \quad
    \avg{xy}_\vxi=\mathcal{R}.
  \end{gathered}
\end{equation}
The averages $\avg{\cdot}_\vxi$ can be performed analytically for all three
learning algorithms and their particular choice of $\mathcal{F}$, see
Tab.~\ref{tab:rules}. We give the results in
Appendix~\ref{sec:dynamics-sde}. This procedure eventually yields a set of
closed equations for $\mathcal{R}$ and $\mathcal{Q}$ for each learning
algorithm, which can be solved numerically.

\begin{figure}
  \centering
  \includegraphics[width=\columnwidth]{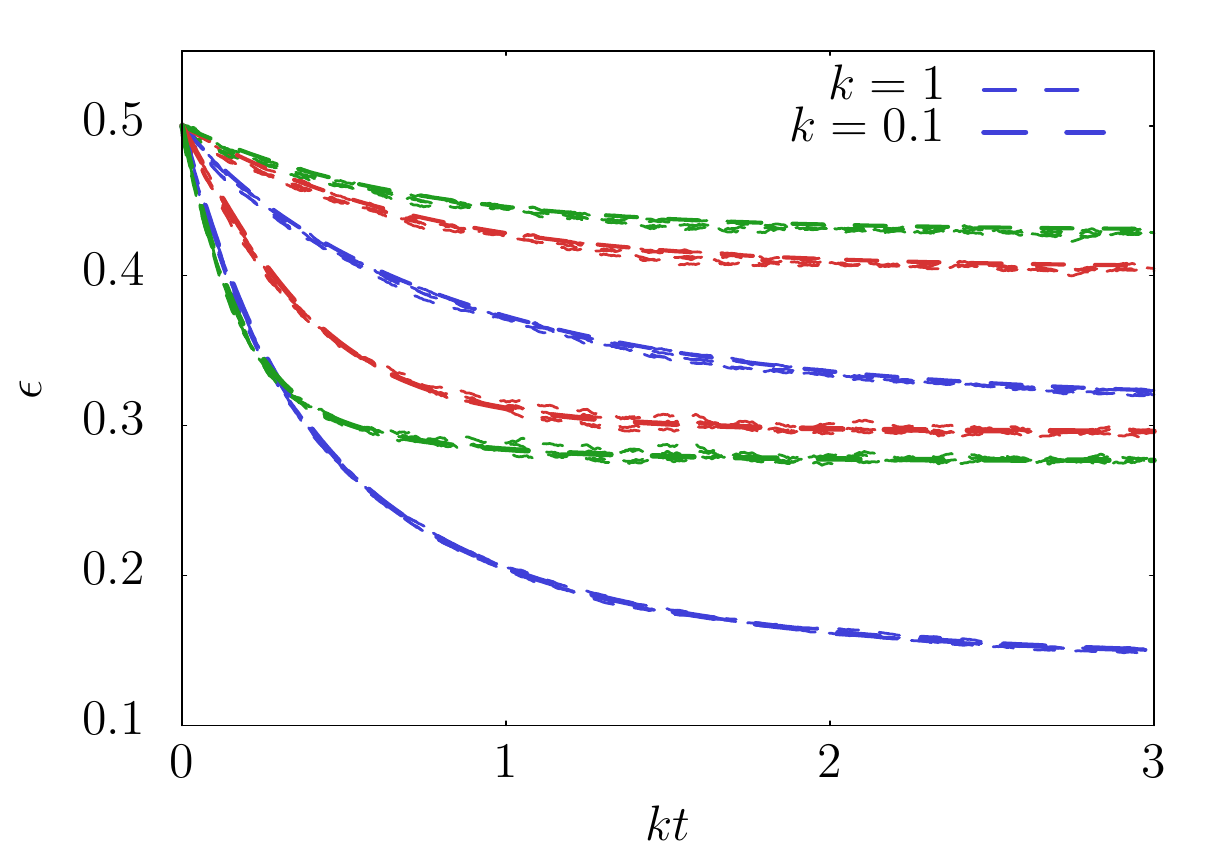}
  \caption{\label{fig:dynamics-thermo}\textbf{Dynamics of online learning in
      large networks.} We compare the generalisation error $\epsilon$ of online
    learning in the thermodynamic limit with fixed normalised learning rate
    $\lrn_0~=~1$ computed by solving the equations of motion for $\mathcal{Q}$
    and $\mathcal{R}$, Eq.~\eqref{eq:eom} (solid) with the results of numerical
    simulations of the noisy dynamics of a network with $N~=~10000$ (dashed)
    performing online learning using the Hebbian~\square{hebbian},
    Perceptron~\square{perceptron} and AdaTron~\square{adatron} algorithms,
    plotting five trajectories for each algorithm. Parameters: $\beta=D=1$.}
\end{figure}

Fig.~\ref{fig:dynamics-thermo} shows the generalisation error obtained from
numerical simulations of the Langevin equation~\eqref{eq:langevin} for a network
with $N=10000$, $\lrn_0=\beta=D=1$ and compares it to the result obtained by our
analytical calculation that we just discussed. First, we note that $\epsilon$ is
a self-averaging quantity, \emph{i.e.} each simulation run generates the same
$\epsilon$ over time within small fluctuations which are scale inversely with
$N$. Furthermore, the dynamics of $\epsilon$ are well described by our
analytical result. While the Hebbian learning takes the longest time to
converge, it is perhaps surprisingly the most robust algorithm in the presence
of noise, consistently yielding the lowest generalisation errors. Indeed, for
online learning with $k=0$ and no noise, it is well established that $\epsilon$
decays slower for the Perceptron than the Hebbian algorithm; on the other hand,
Hebbian learning fails miserably with non-uniform input
distributions~\cite{engel2001}. The performance of the Perceptron is
significantly improved by a choice of time-dependent learning rates in a process
called \emph{annealing}. This is beyond the scope of this paper, but
see~\cite{convergence1995} for a detailed discussion of the impact of
time-dependent learning rates on the convergence of learning algorithms.

\begin{figure}
  \centering
  \includegraphics[width=\columnwidth]{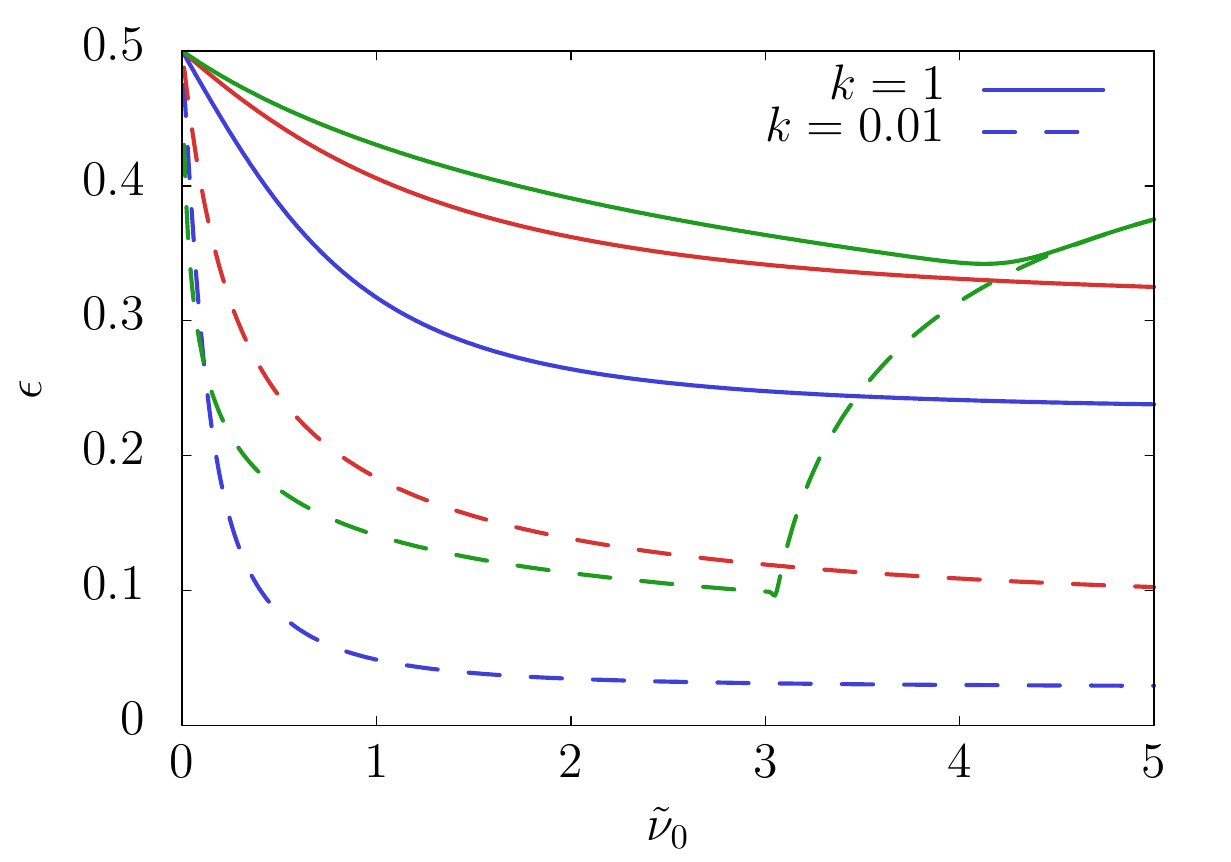}
  \caption{\label{fig:error-thermo-vs-nu}\textbf{Final generalisation error in
      large networks.} We plot the final, steady-state generalisation error
    $\epsilon$ of online learning in the thermodynamic limit using
    Hebbian~\square{hebbian}, Perceptron~\square{perceptron} and
    AdaTron~\square{adatron} algorithms. The behaviour of the algorithms and
    AdaTron learning in particular is discussed in detail in
    Sec.~\ref{sec:examples2-dynamics}. Parameters: $\beta=D=1$.}
\end{figure}

A remarkable property of AdaTron learning is demonstrated in
Fig.~\ref{fig:error-thermo-vs-nu}, where we plot the final, steady-state
generalisation error $\epsilon$ against the normalised learning rate
$\lrn_0$. While Hebbian and Perceptron learning (green and blue, resp.) show
the expected decrease of $\epsilon$ with $\lrn_0$, there is a sharp increase of
$\epsilon$ for AdaTron learning at $\lrn_c=3$ (green). Indeed, for large
learning rates, the AdaTron algorithm will fail completely. This sensitivity of
the algorithm to the value of the learning rate is well-known in the noise-free
case without potential $V(\vw)$~\cite{engel2001} and persists in our model with
noise, most markedly for low potential stiffness $k$. A detailed analysis of the
first-order system $(\mathcal{Q}, \mathcal{R})$ reveals that the fixed point
with $\epsilon$ close to $1$ becomes unstable as the learning rate crosses the
point $\lrn_c=3$ and a second, stable fixed point emerges with $\epsilon\to1/2$
(see Appendix~\ref{sec:adatron-critical-learning-rate}).

\subsection{Efficiency of learning}

\begin{figure*}[ht!]
  \centering
  \includegraphics[width=1.4\columnwidth]{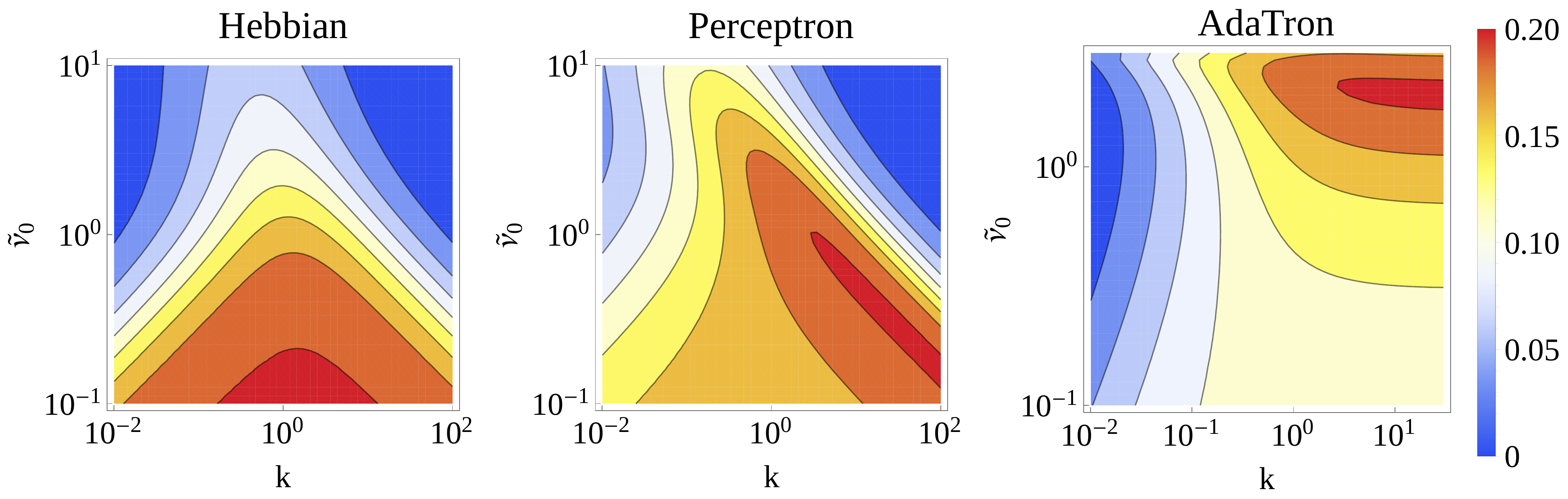}%
  \quad
  \includegraphics[width=.6\columnwidth]{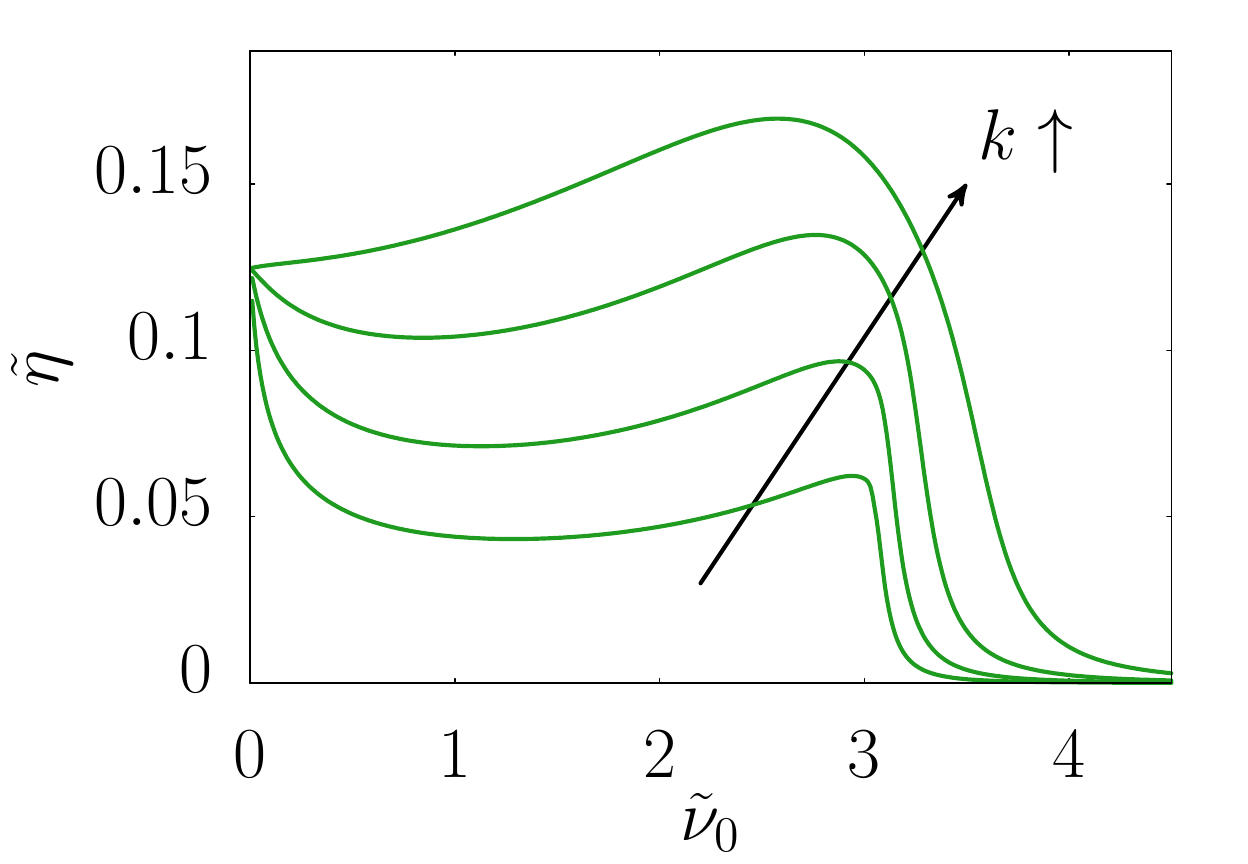}
  \caption{\label{fig:efficiency-thermo}\textbf{Efficiency of learning in large
      networks.} The efficiency $\tilde{\eta}$, Eq.~\eqref{eq:efficiency2}, for
    neural networks performing online learning with fixed normalised learning
    rate~$\lrn_0$ using the Hebbian, Perceptron and AdaTron algorithms for
    online learning in the thermodynamic limit is shown from left to right as a
    function of the potential stiffness~$k$ and~$\lrn_0$. On the far right, we
    plot the same efficiency~$\tilde{\eta}$ of AdaTron learning~\square{adatron}
    versus the normalised learning rate $\lrn_0$ for increasing values of~$k$
    from~$0.01$ (bottom) to~$1$ (top). Parameters: $\beta=D=1$.}
\end{figure*}

We can also derive an ordinary differential equation for the ensemble average of
the excess heat~\eqref{eq:Sex} in terms of $\mathcal{Q}$ and $\mathcal{R}$, with
the details to be found in Appendix~\ref{sec:excess-heat}. Since the components
of the teacher and the weights are normally distributed, the change in Shannon
entropy of the marginalised distribution of a weight $\Delta S(\w_n)$ can be
expressed in terms of just $\mathcal{Q}$, giving us all the information
necessary to compute the efficiency of learning
$\tilde{\eta}$~\eqref{eq:efficiency2}. We plot the efficiency $\tilde{\eta}$ in
the thermodynamic limit in Fig.~\ref{fig:efficiency-thermo} against the
normalised learning rate $\lrn_0$ and the potential stiffness $k$, which are the
only remaining free parameters in this model.

The efficiency of Hebbian learning is roughly symmetric with respect to $k$
around $k=1$, while Perceptron and AdaTron learning display highly asymmetric
patterns. However, we find that despite the different patterns, the maximum
efficiency for all three algorithms is $\eta \simeq 0.2$. We can dig a little
deeper by first noting that since $p(T_n, \w_n)$ is normally distributed for the
learning algorithms we have considered, both the mutual information
$\mutual{T_n}{\w_n}$ and the mutual information between the true and the
predicted label for an arbitrary input $\mutual{\tlab}{\lab}$ can be written as
functions of only the correlation between $T_n$ and $\w_n$,
$\rho\equiv\mathcal{R}/\sqrt{\mathcal{Q}}$. Expanding around $\rho=0$ yields
\begin{equation}
  \frac{\mutual{\tlab}{\lab}}{\mutual{T_n}{\w_n}} = \frac{\ln 2 -
    S(\arccos(\rho)/\pi)}{-1/2 \ln(1-\rho^2)} = \frac{4}{\pi^2} +
  \mathcal{O}(\rho^2) \simeq 0.4
\end{equation}
which turns out to be a good approximation for $\rho\lesssim 0.9$. So at maximum
efficiency,
\begin{equation}
  \frac{\mutual{\w_n}{T_n}}{\Delta S(\w_n)+\Delta Q^\ex_n} \simeq \frac{1}{2}
\end{equation}
for all three algorithms.

The last plot in Fig.~\ref{fig:efficiency-thermo} shows
that the bifurcation we discussed for the AdaTron learning in
Sec.~\ref{sec:examples2-dynamics} leads to a decaying efficiency
$\tilde{\eta}\to0$ since $\mutual{\tlab}{\lab}\to0$. This effect is smoothed out
with increasing potential stiffness.

Let us finally note that, in this model, the rate of heat dissipation of a
single weight diverges in the thermodynamic limit, $\dot{Q}_n\to\infty$ as
$N\to\infty$. This is readily understood from a physical point of view since the
weights experience a large force, $f\sim\sqrt{N}$, which fluctuates very
quickly. This observation reinforces the importance of our second bound
involving the excess heat~\eqref{eq:Sex}, which does not diverge even in the
limit $N\to\infty$.



\section{Concluding perspectives}
\label{sec:conclusion}

We have analysed the learning of linearly separable rules by neural networks as
a model for the thermodynamics of generalisation. Using stochastic
thermodynamics and information theory, we have shown that the accuracy with
which the neuron is able to apply the rule to previously unseen inputs is
constrained by the dissipation of free energy of a single weight during the
learning process. Our results hold for all learning algorithms that have access
to all the weights, $\vw$, and either a set of or a succession of samples
$(\tlab^{\mu(t)}, \vxi^{\mu(t)})$ in batch or online learning, respectively. We
have furthermore given a detailed analysis of both the dynamics and the
thermodynamics of online learning in large neural networks with noisy dynamics
and weights constrained by an external potential.

It is worthwhile to revisit the results of our earlier work~\cite{goldt2017} in
the light of these results. In this previous paper, we studied a different
learning problem, namely the learning of $P$ mappings $\vxi^\mu\to\tlab^\mu$
from \emph{fixed} inputs $\vxi^\mu$, $\mu=1,\ldots,P$ to their true labels
$\tlab^\mu$. The true labels were drawn at random for each input, and hence
uncorrelated to the inputs and to each other. Hence there is no generalisation
error for this problem -- if the true label of every 
input is determined by pure chance, the mappings
$\set{\vxi^\mu\to\tlab^\mu}_{\mu=1}^P$ carry no information about the label of a
previously unseen input. Instead, the challenge is to find a set of weights that
reproduce the mappings faithfully.

The two problems are however related in the following way. It is possible to at
least construct a teacher $\vT$ that reproduces all the mappings
$\vxi^\mu\to\tlab^\mu$ using $\tlab^\mu=\sgn(\vT\cdot\vxi^\mu)$ if and only if
the number of mappings $P$ is less than the capacity of the network. This
capacity is usually defined in the thermodynamic limit, where the number of
weights $N\to\infty$, and we are interested in the relative number of inputs
$\alpha_c\equiv P_c/N\sim 1$ for which there exists a teacher $\vT$ that
reproduces all the true labels via $\tlab=\sgn(\vT\cdot\vxi)$ with probability
1~\cite{mackay2003}. Its numerical value can be derived analytically from
replica calculations~\cite{gardner1987}, but it was first understood using
geometrical arguments~\cite{polya1954,cover1965} (see also~\cite{mackay2003} for
a pedagogical discussion).

If it is possible to construct a teacher $\vT$, the rule implicitly defined by
the mappings is realisable and can, at least in theory, be learned. Even in that
case, however, the issue remains for the scenario considered in~\cite{goldt2017}
that the number of samples from which the neuron learns is limited and might not
be sufficient to learn the underlying ``rule'' effectively. On the other hand,
learning the mappings $\set{\vxi^\mu\to\tlab[\mu]}_\mu^P$ is still a meaningful
task even if it is not possible to even construct a network that reproduces them
all, if one is willing to accept a certain error in the predictions of the
network.

Neurons with a single-layer architecture described here are generally limited to
the implementation of linearly separable Boolean functions $\vxi\to\pm1$. A
natural generalisation is to consider networks with several layers, where the
output of the neurons in one layer is the input for the neurons in the next
layer~\cite{engel2001,mackay2003}. The capabilities of networks with
intermediate layers are remarkable: a network of binary neurons ($\lab=\pm1$)
with just a single intermediate layer can implement \emph{any} Boolean function
of its inputs~\cite{denker1987}, while a network with continuous neurons
($\lab\in\mathbb{R}$, \emph{e.g.} $\lab=\tanh(\act)$) is capable of
approximating any function of its inputs to any required
accuracy~\cite{debao1993}\footnote{In both cases, the intermediate layer
  requires a sufficient number of neurons}. However, the analysis of these
networks is much more involved than that of the single-layer feedforward network
discussed here and is left for future work.

Besides the generalisation to networks with intermediate layers of neurons, this
work opens up numerous other avenues for further research. It would be
intriguing to consider the generalisation of our model to multi-valued teacher
functions, \emph{e.g.} for a network learning to classify digits. The teacher
could also be made subject to noise in its outputs $\tlab$, or its components,
$T_n$, or both. Another intriguing learning problem is that of a changing
environment, modeled by a drifting
teacher~\cite{kinouchi1993,biehl1993a}. Designing a learning algorithm that
optimises the thermodynamic efficiency looks like a serious challenge. More
broadly, studying the thermodynamic costs of learning to generalise might form a
suitable basis to consider the thermodynamics of
decision-making~\cite{colabrese2017}.

We have considered the limitations on computation in neural networks that are a
consequence of the second law of thermodynamics and account for the free energy
costs of computations. However, there are further limiting factors on the
ability to compute. One is the availability of data: for a given inference
problem, such as inferring a teacher $\vT$ from a number of samples
$(\tlab, \vxi)$, there is a minimum amount of data that is required to make any
prediction that is better than simply flipping a coin. A second limiting factor
is time: ideally, we would like to have an algorithm whose completion time
scales polynomially with the system size, rather than exponentially. It has
recently become clear that these two constraints can be understood using
statistical mechanics in terms of phase transitions~\cite{zdeborova2016}. It
will be interesting to see whether the thermodynamic limits that we have derived
in this paper fit into this picture, and if so, where they can be found.

\begin{acknowledgments}
  We thank David Hartich for valuable discussions and careful reading of the
  manuscript and Hans-Günther Döbereiner for stimulating discussions of the
  science of decisions.
\end{acknowledgments}

\appendix

\section*{Appendices}

\noindent The following appendices give a detailed proof of our main results,
inequalities~\eqref{eq:inequality1} and~\eqref{eq:inequality2}, in
Appendices~\ref{sec:derivation1}
to~\ref{sec:derivation2}. Appendix~\ref{sec:batch-learning} discusses how our
results apply to batch learning. Detailed calculations for the learning dynamics
in the thermodynamic limit are given in Appendices~\ref{sec:dynamics-sde}
to~\ref{sec:excess-heat}.

\section{Derivation of inequality~\eqref{eq:inequality1}}
\label{sec:derivation1}

Our first main result, Eq.~\eqref{eq:inequality1}, can be derived from the
second law of stochastic thermodynamics~\cite{seifert2012} which states that the
rate of total entropy production of the full system is positive
\begin{equation}
  \dot{S}^\tot(t) \ge 0.
\end{equation}
We will drop the explicit time argument in the following discussion but
emphasise that since the distribution $p(\vT, \vw, t)$ is time-dependent, so are
of course all the quantities derived from it.

We discussed in Sec.~\ref{sec:dynamics} that the total probability current of
the system decomposes into a separate current due to the fluctuations in every
subsystem $\w_n$, see Eq.~\eqref{eq:fpe}. As a consequence, the rate of total
entropy production $\dot{S}^\tot$ of the network can be split into separate,
non-negative contributions due to the fluctuations of every subsystem $\w_n$,
$\dot{S}^\tot_n$, each of which can further be split into two separate
contributions, so that we have
\begin{equation}
  \dot{S}^\tot = \sum_n \dot{S}^\tot_n = \sum_n \dot{S}_n(\vw, \vT) + \sum_n
  \dot{S}^\m_n \ge0.
\end{equation}
The first part is the rate of change of the Shannon entropy of the full
distribution $p(\vT, \vw, t)$ due to the dynamics of $\w_n$ and reads
\begin{equation}
  \dot{S}_n(\vT, \vw) = - \int \dd \vT \dd \vw \;  j_n(\vT, \vw)
  \partial_n \ln p(\vT, \vw).
\end{equation}
while the rate of thermodynamic entropy production in the medium by the $n$th
weight is given by
\begin{equation}
  \dot{S}^\m_n =\int \dd \vT \dd \vw \;  j_n(\vT, \vw) F_n(\vT, \vw)
\end{equation}
for with the temperature set to unity.

Writing $p(\vT, \vw) = p(\w_n)p(\vT, \others{\vw}_n|\w_n)$ where 
$\others{\vw}_n\equiv(\w_1, \dots, \w_{n-1}, \w_{n+1}, \dots, \w_N)$, we have
\begin{equation}
  \label{eq:S_n-splitted}
  \dot{S}_n(\vT, \vw) = \partial_t S(\w_n) - l_n(\w_n:\vT, \others{\vw}_n)
\end{equation}
where $\partial_t S(\w_n)$ is the change of the Shannon
entropy~\eqref{eq:shannon} of the marginalised distribution
$p(\w_n)=\int \dd \vT \dd \others{\vw_n} \; p(\vT, \vw)$ due to the dynamics of
the $n$th weight. Throughout this paper, we use the dot notation to denote
time-dependent rates, as opposed to the time derivative of state functions like
the Shannon entropy. The second term in Eq.~\eqref{eq:S_n-splitted} is an
information-theoretic piece which gives the rate at which the dynamics of $\w_n$
changes its correlations with the other degrees of freedom in the system, namely
the other weights $\others{\vw}_n$ and the teacher $\vT$, as measured by the
mutual information $\mutual{\w_n}{\vT, \others{\vw}_n}$. This quantity has been
introduced as the (thermodynamic) learning
rate~\cite{allahverdyan2009,hartich2014}, not to be confused with the learning
rate $\lr(t)$ introduced in Eq.~\eqref{eq:learning-force}, or the ``information
flow''~\cite{horowitz2014,horowitz2015}. Its explicit form is
\begin{multline}
  \label{eq:learning-rate}
  l_n(\w_n : \vT, \others{\vw}_n) =\\ \int \dd \vT \dd \vw \; j_n(\vT, \vw, t) \partial_n \ln p(\vT, \others{\vw}_n|\w_n).
\end{multline}

We finally note that for the isothermal environment that we assume in this
paper, the rate of thermodynamic entropy production is the heat dissipated into
the environment, $\dot{S}^\m_n=\dot{Q}_n$, where we remind ourselves that we have
set the temperature to unity.

Putting it all together, we can formulate the second law for a subsystem,
\begin{equation}
  \label{eq:second-law}
  \dot{S}^\tot_n = \partial_t S(\w_n) + \dot{Q}_n - l_n(\w_n:\vT, \others{\vw}_n) \ge 0.
\end{equation}
which is the starting point of our derivation.

Integrating the $N$ second laws~\eqref{eq:second-law} with respect to time from
$t'=0$ to $t>0$ yields
\begin{equation}
  \label{eq:second-law-integrated}
  \sum_n^N\left[\Delta S(\w_n)+\Delta Q_n\right] \ge \sum_n^{N}\int_0^t \dd t' \; l_n(\w_n : \vT, \others{\vw}_n)
\end{equation}
where we write $\Delta S(\w_n)$ and $\Delta Q_n$ to denote the total change in
Shannon entropy of the distribution $p(\w_n)$ and the total heat dissipated by the
dynamics of the $n$th weight up to time $t$, respectively. We can interpret the
right-hand side of~\eqref{eq:second-law-integrated} by computing the
time-derivative of the mutual information $\mutual{\vT}{\vw}$,
\begin{equation}
  \partial_t \mutual{\vT}{\vw} =
  \int \dd \vT \dd \vw \;\left[ \partial_t p(\vT, \vw, t) \right] \ln
  \frac{p(\vT, \vw, t)}{p(\vT, t)p(\vw, t)}.
\end{equation}
\begin{widetext}
Using the Fokker-Planck Eq.~\eqref{eq:fpe} and integrating by parts, we find
\begin{align}
  \partial_t \mutual{\vT}{\vw} =& \sum_n^N \int \dd \vT \dd \vw \; j_n(\vT, \vw,
                                  t) \partial_n \ln \frac{p(\vT, \vw,
                                  t)}{p(\vT, t)p(\vw, t)}  \\
  =& \sum_n^N \int \dd \vT \dd \vw \; j_n(\vT, \vw,
     t) \partial_n \ln \frac{p(\vT, \vw, t)}{p(\w_n, t)p(\others{\vw}_n|\w_n, t)}  \\
  =& \sum_n^{N} l_n(\w_n : \vT, \others{\vw}_n) - \sum_n^N \int \dd \vT \dd \vw \; j_n(\vT, \vw,
     t) \partial_n \ln p(\others{\vw}_n|\w_n, t) \\
  =& \sum_n^{N} l_n(\w_n : \vT, \others{\vw}_n) - \left( - \partial_t S(\vw) + \sum_n^N \partial_t S(\w_n)\right) \label{eq:last-line}
\end{align}
where in the penultimate line, we have recovered the integrand on the right-hand
side of~\eqref{eq:second-law-integrated}. Integrating the term in brackets in
Eq.~\eqref{eq:last-line} with respect to time yields for all times $t>0$
\begin{equation}
  \int_0^t \dd t' \; \left( \sum_n^N \partial_{t'} S(\w_n) - \partial_{t'}
    S(\vw) \right) =
  \sum_n^N S(\w_n) - S(\vw) \ge 0
\end{equation}
where we have used that at time $t=0$, all the weights are independent of each
other and hence $S(\vw)=\sum_nS(\w_n)$. The inequality follows from the fact
that for any set of random variables,
$\sum_n S(\w_n)\ge S(\vw)$~\cite{cover2006a}.  Using this inequality, we can
deduce from~\eqref{eq:last-line} that
\begin{equation}
  \sum_n^N\left[\Delta S(\w_n) + \Delta Q_n \right]\ge \sum_n^N \int_0^t \dd t' \; l_n(\w_n : \vT, \others{\vw}_n) \ge \mutual{\vT}{\vw}.
\end{equation}
\end{widetext}

Using the chain rule of mutual information~\cite{cover2006a} and the fact that
the components $\vT$ are independent of each other (Sec.~\ref{sec:model}), we
have
\begin{align}
  \mutual{\vT}{\vw} =& \sum_n^N I(T_n:\vw|T_{n-1},\dots,T_1)\\
  \begin{split}
    =&\sum_n^N \mutual{T_n}{\vw, T_{n-1},\dots,T_1} -\\
    & \qquad \sum_n^N \mutual{T_n}{T_{n-1},\dots,T_1}
  \end{split}\\
  \ge & \sum_n^N\mutual{T_n}{\vw}\\
  = & \sum_n^N \mutual{T_n}{\w_n} + \mutualGiven{T_n}{\others{\vw}_n}{\w_n}\\
  \ge & N\mutual{T_n}{\w_n}
\end{align}
where the inequalities follow again from the non-negativity of mutual
information and the fact that all the weights and all the components of the
teacher are statistically identical. Using the latter argument for the total
entropy production and inserting our last result into the integrated form of the
second law~\eqref{eq:second-law-integrated}, we find that
\begin{equation}
  \label{eq:second-law-nth}
  \Delta S(\w_n)+\Delta Q_n \ge \mutual{\w_n}{T_n}
\end{equation}

Finally, we need to show that the mutual information between the $n$th component
of the weight and teacher vectors are an upper bound on the mutual information
between the true and predicted labels of any sample $\vxi$,
\begin{equation}
  \label{eq:mutual-info-inequality}
  \mutual{\w_n}{T_n}\ge\mutual{\tlab}{\lab}.
\end{equation}
Our strategy will be to show that the
inequality~\eqref{eq:mutual-info-inequality} holds even if the neuron predicts a
label deterministically, $\lab=\sgn(\vw\cdot\vxi)$. The generalisation error is
then the lowest for a given noise level in the weights and given by
$\mutual{\tlab}{\lab}=\ln 2 - S(\epsilon)$.

Let us first consider a network with $N=1$. We start by noting that for
arbitrary random variables $X$ and $Y$ and an arbitrary function $F(Y)$, we can
always write $p(x, y, f(y))=p(x)p(y|x)p\left(f(y)|y\right)$. We thus identify
$X\to Y\to F$ as a Markov chain and find
\begin{equation}
  \label{eq:dpi}
  I(X:Y) \ge I(X:F)
\end{equation}
using the data processing inequality~\cite{cover2006a}. For $N=1$, we can apply
this result twice to show that
\begin{equation}
  \label{eq:2}
  \mutual{\w}{T} \ge \mutual{\w \xi}{T\xi} \ge \mutual{\lab}{\tlab}
\end{equation}
as required.


In the thermodynamic limit $N\to\infty$, we use the auxiliary variables
$x\equiv \vw\cdot\vxi/\sqrt{N}$ and $y\equiv\vT\cdot\vxi/\sqrt{N}$. We then have
from~\eqref{eq:dpi}
\begin{equation}
  \mutual{\tlab}{\lab} \le \mutual{x}{y}
\end{equation}
since $\tlab$ and $\lab$ are functions of $x$ and $y$, Eq.~\eqref{eq:tlab} and
Eq.~\eqref{eq:lab}, respectively. We can now average $x$ and $y$ over the
inputs~\eqref{eq:inputs} using $\avg{\xi_n}_\xi=0$ and
$\avg{\xi_n\xi_m}_\xi=\delta_{nm}$. By the central limit theorem, $x$ and $y$
are then distributed according to a bivariate Gaussian distribution with
correlation~\cite{cover2006a}
\begin{equation}
  \label{eq:correlation}
  \rho \equiv \frac{\text{cov}(\w_n, T_n)}{\text{sd}(\w_n) \, \text{sd}(T_n)} =
  \frac{\vw\cdot\vT}{|\vw||\vT|}
\end{equation}
This is a crucial step in our derivation since it allows us to connect the
statistics of teacher and weight in one dimension to the statistics of the true
and predicted labels, which are functions of the vectors $\vT$ and $\vw$. The
mutual information of two variables with a bivariate Gaussian distribution is a
function of their correlation alone~\cite{cover2006a},
\begin{equation}
  \label{eq:mutual-gaussian}
   I_G(\w_n:T_n)= -\frac{1}{2}\left( 1-\ln \rho^2 \right) = \mutual{x}{y} 
\end{equation}
which would \emph{also} be the mutual information $\mutual{\w_n}{T_n}$ if $\w_n$
and $T_n$ were jointly distributed normally, which they are not
necessarily. However, we can show that $I_G(\w_n:T_n)$ is a lower bound on
$\mutual{\w_n}{T_n}$ using the maximum entropy principle. This is a prescription
for finding the probability distribution that maximises the Shannon entropy
given a number of constraint on the distribution, usually in the form of fixed
moments. We briefly review this concept in Appendix~\ref{sec:max-ent}. The
crucial point here is that a Gaussian distribution is the maximum entropy
distribution for a given covariance matrix. We will denote the maximum entropy
notations with an asterisk, \emph{e.g.}~$p^*$.

The mutual information $\mutual{\w_n}{T_n}$ can be expressed as the relative
entropy or Kullback-Leibler divergence between the joint distribution
$p(\w_n, T_n)$ and the factorised distribution $p(T_n)p(\w_n)$~\cite{cover2006a}:
\begin{align}
  \label{eq:kl-distance}
  \mutual{\w_n}{T_n} =& \kl{p(T_n, \w_n)}{p(T_n)p(\w_n)}\\
               \equiv&  \int \dd \vT \dd \vw \; p(\vT, \vw) \ln
                       \frac{p(\vT,\vw)}{p(\vT)p(\vw)}
\end{align}
where the inequality is true for arbitrary probability
distributions. Introducing the shorthand $p(\w_n)\equiv p_\w$ etc.\ to simplify
notation, we hence are left to show that
\begin{align*}
  & \mutual{\w_n}{T_n} - I_G(\w_n:T_n) \\[1em]
  = & \avg{\ln \frac{p_{T\w}}{p_T p_\w}}_p -   \avg{\ln \frac{p^*_{T\w}}{p^*_T p^*_\w}}_{p^*} \\[1em]
  = & \avg{\ln p_{T\w}}_p - \avg{\ln p^*_{T\w}}_{p*}\\[.5em]
  & \quad - \left[ \avg{\ln p_T}_p - \avg{\ln p^*_T}_{p^*} + \avg{\ln p_\w}_p -
    \avg{\ln p^*_\w}_{p^*}\right]\\[1em]
  = & \avg{\ln p_{T\w}}_p - \avg{\ln p^*_{T\w}}_p\\[.5em]
  & \quad - \left[ \avg{\ln p_T}_p - \avg{\ln p^*_T}_p + \avg{\ln p_\w}_p -
    \avg{\ln p^*_\w}_p\right]\\[1em]
  = & \kl{p_{T\w}}{p^*_{T\w}}-\kl{p_{T}}{p^*_{T}}-\kl{p_{\w}}{p^*_{\w}}\\[1em]
  = & \kl{p_{T|\w}}{p^*_{T|\w}} \ge 0
\end{align*}
where we used that $\avg{\ln p^*_{T\w}}_{p^*} =\avg{\ln p^*_{T\w}}_p$ for the
third equality, see Sec.~\ref{sec:max-ent}, while for the last equality we
applied the chain rule for the Kullback-Leibler distance~\cite{cover2006a} and
remembered that $p(T_n)$ is a Gaussian distribution and hence the maximum
entropy distribution for a given variance. This completes our derivation of the
bound~\eqref{eq:inequality1}.

\section{Surprise and maximum entropy distributions}
\label{sec:max-ent}

We briefly review the concept of a maximum entropy distributions, which have a
long history in physics~\cite{jaynes1957a}. We will focus on the case of a
single variable to illustrate the concepts, but we note that the
multi-dimensional case can be treated using the same methods.

We are looking for a probability distribution $p$ of a continuous variable $X$
with support $\mathcal{S}$ which is subject to $M$ constraints, namely
\begin{equation}
  \label{eq:constraints}
  \int_\mathcal{S} \dd x \; p(x)r_i(x) = \alpha_i.
\end{equation}
The maximum entropy prescription for finding the distribution $p(x)$ is to find
the distribution that maximises the Shannon entropy $S(X)$~\eqref{eq:shannon}
under the constraints~\eqref{eq:constraints}. This is a standard calculation
using variational calculus; here we simply quote the result~\cite{cover2006a},
\begin{equation}
  \label{eq:p-max-ent}
  p^*(x) \sim \exp \left( \lambda_0 - 1 + \sum_i^M \lambda_i r_i(x) \right)
\end{equation}
where $\lambda_i$ are the Lagrange multipliers chosen such that $p(x)$ obeys the
constraints. Proving that~\eqref{eq:p-max-ent} is a maximum is a rather involved
calculation and is more easily proven using information-theoretic
methods~\cite{cover2006a}.

The key point for our purposes is that for a distribution $p(x)$, which is
unknown except for a number of its moments, averaging the surprise $\ln p^*(x)$
of the maximum entropy distribution for the known moments over $p^*$ is equal to
the average taken with respect to the true distribution $p$,
\begin{equation}
  \avg{\ln p^*(x)}_{p^*}  =\avg{\ln p^*(x)}_p.
\end{equation}
This result is a direct consequence of the form of the maximum entropy
distribution and the fact that the moments that enter $p^*$ are by construction
equal to the corresponding moments of $p$.

\section{Derivation of inequality~\eqref{eq:inequality2}}
\label{sec:derivation2}

The non-adiabatic entropy production of the $n$th single weight is defined as
\begin{equation}
  \label{eq:Sna-n}
  \dot{S}^\na_n(t) \equiv \int \dd \vT \dd \vw \;
  p \left(\frac{j_n}{p} - \frac{j_n^\st}{p^\st} \right)^2 \ge 0,
\end{equation}
see Sec.~\ref{sec:results2} for a detailed discussion.  Summing
$\dot{S}^\na_n(t)$ over all subsystems, we find
\begin{equation}
  \sum_n \dot{S}^\na_n(t) = \sum_n \dot{S}_n(\vT, \vw) + \sum_n \dot{Q}^\ex_n
  \ge 0 
\end{equation}
where the rate of excess heat production $\dot{Q}^\ex_n$ was defined in
Eq.~\eqref{eq:Sex}. After writing
$\dot{S}_n(\vT, \vw)=\partial_t S(\w_n) - l_n(\w_n : \vT, \others{\vw}_n)$ and
integrating over time, see the discussion in Appendix~\ref{sec:derivation1}, we
find that
\begin{equation}
  \label{eq:Sna-integrated}
  \sum_n^N\left[\Delta S(\w_n)+\Delta Q^\ex_n\right] \ge \sum_n^{N}\int_0^t \dd t' \; l_n(\w_n : \vT, \others{\vw}_n).
\end{equation}
and we can now proceed along the lines of Appendix~\ref{sec:derivation1}.

\section{Batch learning}
\label{sec:batch-learning}

Our discussion has focused on online learning, where, at any one point in time,
the network experiences a learning force $\vec{f}$ due to a single input and its
label, Eq.~\eqref{eq:learning-force}. Another approach is to average the
learning force over a set $D=\set{(\tlab[\mu], \vxi^\mu)}_{\mu=1}^P$ of $P$
inputs and their labels,
\begin{equation}
  f_n \equiv \lr(t) \avg{\xi_n^\mu \tlab^\mu \mathcal{F}(|\vw(t)|, \vw(t)\cdot\vxi^\mu, \vT\cdot\vxi^\mu)}_D.
\end{equation}
This strategy is called \emph{batch learning} and $P$ is usually chosen to be on
the order of $N$. In the thermodynamic limit, as $N\to\infty$ one thus lets
$P\to\infty$ while keeping the ratio $\alpha\equiv P/N$ on the order of one.

Batch learning clearly comes with high requirements in terms of memory. It is
generally more efficient than online learning, although the latter can achieve
generalisation errors which at least asymptotically match the results from batch
learning~\cite{engel2001}.

Our two main results, inequalities~\eqref{eq:inequality1}
and~\eqref{eq:inequality2} apply to batch learning as well. This is because in
our derivation, we only used the fact that the teacher enters the force on the
weights, albeit indirectly. We did not have to specify the exact form of the
learning force that introduces the correlations between the weight and the
teacher. Hence it does not make a difference in the derivation of the
inequalities whether the learning force is computed for just a single sample or
averaged over a set of samples.

\begin{widetext}
\section{Solving the learning dynamics in the thermodynamic limit}
\label{sec:dynamics-sde}

Here we give a detailed derivation of the equations of motion for the order
parameters $\mathcal{Q}$ and $\mathcal{R}$ introduced in
Sec.~\ref{sec:examples2} in the thermodynamic limit $N\to\infty$. These are most
easily derived by rewriting the Langevin equations for the weights,
Eq.~\eqref{eq:langevin}, as Itô stochastic differential
equations~\cite{gardiner2009}
\begin{equation}
  \label{eq:ito}
  \dd \vw(t) = -k \vw(t) \dd t
  + \lr(t) \vxi^{\mu(t)} \tlab^{\mu(t)} \mathcal{F}(|\vw(t)|,
  \vw(t)\cdot\vxi^{\mu(t)}, \vT\cdot\vxi^{\mu(t)})\dd t + \dd \vec{W}(t).
\end{equation}
The random Wiener process has components $\dd W_n(t)$ which are normally
distributed with mean 0 and variance $2 D\dd t=2 \dd t$ in our choice of
units. It is related to the noise term of the Langevin equation via
$\dd W_n(t)=\int_t^{t+\dd t}\dd t'\zeta_n(t')$; see~\cite{gardiner2009} for more
details. All other symbols take the same meaning as discussed before
Eq.~\eqref{eq:langevin}. We assume that the inputs that enter the equation are
changing on a timescale much faster than the relaxation time of the
weights. Hence it is only the statistical properties of the inputs that
determine the dynamics of $\vw$ in the thermodynamic limit. We can thus average
over the inputs, making the detailed dynamics of $\mu(t)$ unimportant. We can
derive the equations of motion for the means of
$\mathcal{Q}\equiv \vw\cdot\vw/N$ and $\mathcal{R}\equiv\vT\cdot\vw/N$ by
expanding to \emph{second} order in $\dd \vw$ and keeping terms on the order of
$\dd t$:
\begin{equation}
  \label{eq:dQ}
  \begin{aligned}
    \dd \mathcal{Q} \equiv& \mathcal{Q}(\vw + \dd \vw) - \mathcal{Q}(\vw) \\
    =&\frac{1}{N}\left(2 \vw \cdot \dd \vw + \dd \vw \cdot \dd \vw\right) \\
    =& 2 (1 - k \mathcal{Q}) \dd t + \lrn(t)^2 \avg{\mathcal{F}(\sqrt{\mathcal{Q}},
      \vT\cdot\vxi, \vw\cdot\vxi)^2} \dd t + 2 \lrn(t) \avg{\sgn (\vT \cdot \vxi)
      \frac{\vw \cdot \vxi}{\sqrt{N}} \mathcal{F}(\sqrt{\mathcal{Q}},
      \vT\cdot\vxi, \vw\cdot\vxi)} \dd t
  \end{aligned}
\end{equation}
where, contrary to ordinary calculus, the term $\dd \vw\cdot\dd \vw$ has
contributed two terms, one from the Wiener process and one because
$\vxi\cdot\vxi\approx N\approx 1/\dd t$. We have also used the scaling
$\lr(t) = \sqrt{N}\lrn(t)$ that we discussed in Sec.~\ref{sec:scal-learn-rate}
of the main text. In this section, we will denote by $\avg{\cdot}$ the average
with respect to the distribution of inputs~\eqref{eq:inputs}. Likewise, we have
\begin{align}
  \label{eq:dR}
  \dd \mathcal{R} =& \vT \cdot \dd \vw / N \nonumber \\
                  =& -k \mathcal{R} \dd t + \lrn(t) \sgn (\vT \cdot \vxi) \frac{\vT
                     \cdot \vxi}{\sqrt{N}} \mathcal{F} \dd t
\end{align}
We can simplify the averages over the inputs and thus these equations by noting
that the inputs only enter as products
\begin{equation}
  \label{eq:sde_x-and-y}
  x \equiv \frac{\vT\cdot\vxi}{\sqrt{N}} \quad \text{and} \quad y \equiv \frac{\vw\cdot\vxi}{\sqrt{N}},
\end{equation}
such that
\begin{align}
  \dd \mathcal{Q} =& 2(1-k\mathcal{Q}) \dd t +
                     2 \lrn(t)\avg{\sgn(x)y\mathcal{F}(x, y)}\dd t  + \lrn(t)^2\avg{\mathcal{F}(x, y)^2} \dd t\\
  \dd \mathcal{R} =& -k \mathcal{R} \dd t + \lrn(t) \avg{\sgn(x)x\mathcal{F}(x, y)}\dd t
\end{align}
The crucial point to compute the averages $\avg{\cdot}$ is now to realise that
$p(x, y)$ is a binormal Gaussian distribution due to the central limit theorem,
with moments
\begin{gather*}
  \avg{x} = \avg{y} = 0\\
  \avg{x^2} = 1, \avg{y^2} = \mathcal{Q}\\
  \avg{xy} = \frac{1}{N}\sum_{n,m} T_m \w_n \avg{\xi_n\xi_m} =
  \frac{\vT\cdot\vw}{N} = \mathcal{R}
\end{gather*}
where we have used $\avg{\xi_n}=0$ and $\avg{\xi_n\xi_m}=\delta_{nm}$ from
Eq.~\eqref{eq:inputs}. Here we only give the results of these integrals for the
different learning algorithms for completeness, see \emph{e.g.}~\cite{mace1998}
for details on how to perform these integrals~\footnote{N.B. they use a
  different normalisation procedure from us.}.

For Hebbian learning, we have $\mathcal{F}=1$ and hence
\begin{align}
  \label{eq:integrals-hebbian}
  \avg{\sgn(x)y} =& \frac{2}{\pi}\mathcal{R},\\
  \avg{\sgn(x)x} =& \frac{2}{\pi}.
\end{align}
The perceptron algorithm has $\mathcal{F}(x,y)=\theta(-xy)$ such that 
\begin{align}
  \label{eq:integrals-perceptron}
  \avg{\sgn(x)y\mathcal{F}} =& \frac{\mathcal{R}-\sqrt{\mathcal{Q}}}{\sqrt{2 \pi }},\\
  \avg{\sgn(x)x\mathcal{F}} =& \frac{1-\mathcal{R}/\sqrt{\mathcal{Q}}}{\sqrt{2 \pi }},\\
  \avg{\mathcal{F}^2} =& \frac{1}{2}-\frac{1}{\pi}\atan\frac{\mathcal{R}}{\sqrt{\mathcal{Q}-\mathcal{R}^2}}.
\end{align}
Finally, for AdaTron learning with $\mathcal{F}(x, y)=|y|\theta(-xy)$, we find
  \begin{align}
    \avg{\sgn(x)y \mathcal{F}} =& \frac{\mathcal{R} \sqrt{\mathcal{Q}-\mathcal{R}^2}}{\pi }
                                  + \mathcal{Q}\left(- \frac{1}{2}+\frac{1}{\pi}\atan\frac{\mathcal{R}}{\sqrt{\mathcal{Q}-\mathcal{R}^2}}\right),\\
    \avg{\sgn(x)x \mathcal{F}} =& \frac{\sqrt{\mathcal{Q}-\mathcal{R}^2}}{\pi}
                                  + \mathcal{R}\left(- \frac{1}{2}+\frac{1}{\pi}\atan\frac{\mathcal{R}}{\sqrt{\mathcal{Q}-\mathcal{R}^2}}\right),\\
    \avg{\mathcal{F}^2} =& -\frac{\mathcal{R} \sqrt{\mathcal{Q}-\mathcal{R}^2}}{\pi } + \frac{\mathcal{Q}}{2\pi}\left(\pi-
                           2\atan\frac{\mathcal{R}}{\sqrt{\mathcal{Q}-\mathcal{R}^2}}\right).
  \end{align}
Substituting these results into the equations for $\dot{\mathcal{Q}}$ and
$\dot{\mathcal{R}}$, Eq.~\eqref{eq:eom}, yields a closed set of equations in
$\mathcal{Q}$ and $\mathcal{R}$, which can be solved numerically.

\section{Critical learning rate for AdaTron learning}
\label{sec:adatron-critical-learning-rate}

\begin{figure}
  \includegraphics[width=.6\columnwidth]{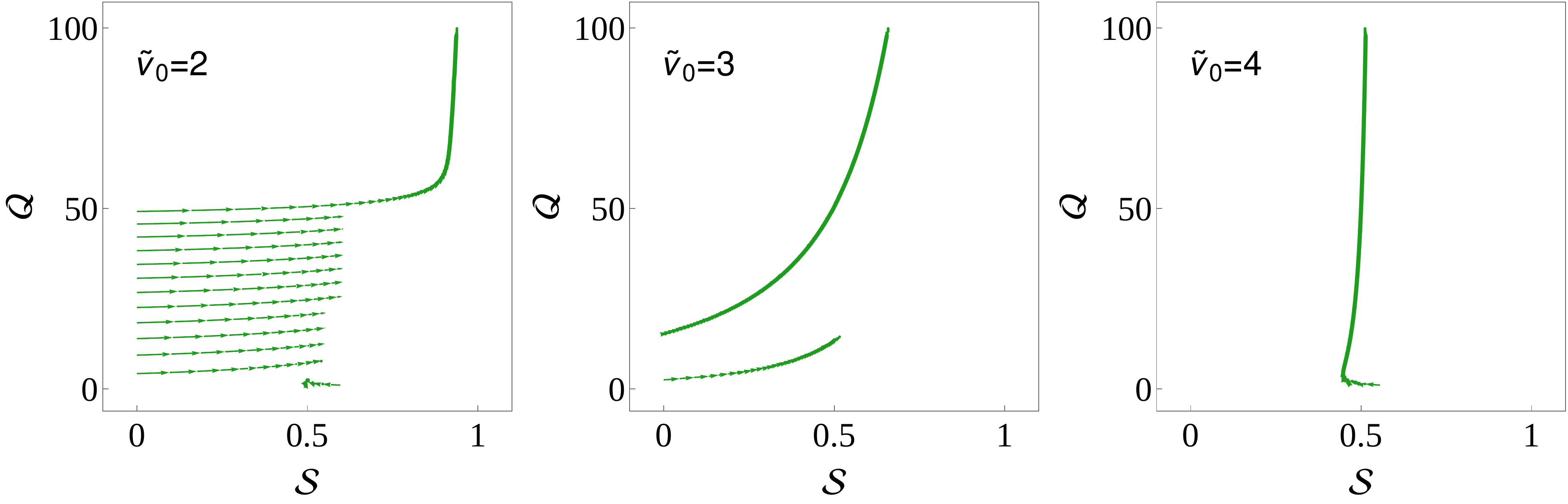}%
  \quad
  \includegraphics[width=.3\columnwidth]{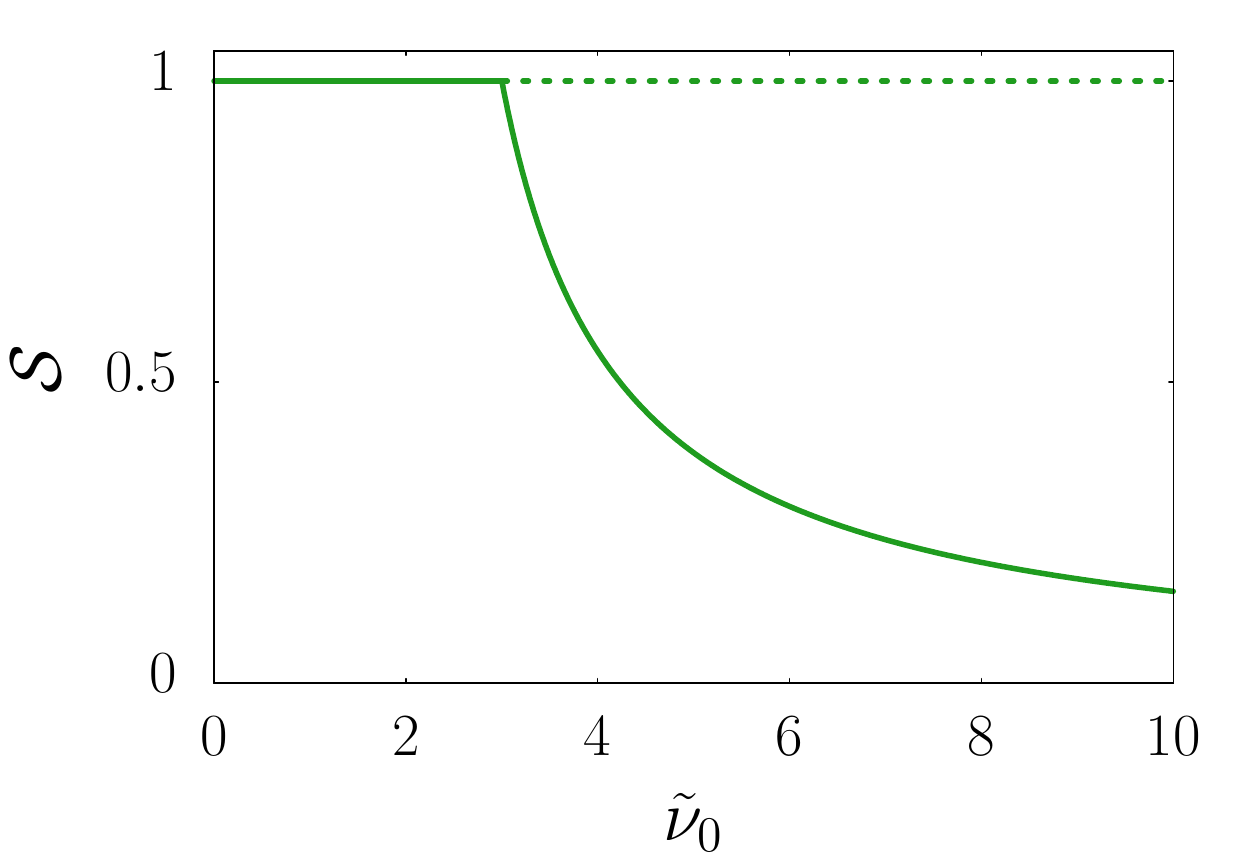}
  \caption{\label{fig:adatron-dynamics}\textbf{Critical learning rate for
      AdaTron learning.} The first three plots from left to right are vector
    plots in phase space for the first-order system
    $(\dot{\mathcal{Q}}, \dot{\mathcal{S}})$, Eq.~\eqref{eq:eom-Q-and-S}, for
    AdaTron learning in the thermodynamic limit $N\to\infty$ with constant
    normalised learning rate $\lrn(t)=\lrn_0$ and $k=0.01$. For $\lrn_0\le2$,
    there is an attracting state with $\mathcal{S}\to1$. As we increase the
    learning rate $\lrn_0$, another attracting state appears with
    $\mathcal{S}\to1/2$ and hence $\epsilon=1/2$. In the limit $k\ll1$, this
    behaviour can be understood from the bifurcation diagram of the closed,
    single equation for $\mathcal{S}$, Eq.~\eqref{eq:eom-S}, shown on the far
    right, where stable (unstable) fixed points are indicated by straight
    (dashed) lines. Parameters: $\beta=D=1$.}
\end{figure}

The critical dependence of the generalisation error $\epsilon$ on the learning
rate $\lr(t)=\sqrt{N}\lrn_0$ for AdaTron learning in weak potentials ($k\ll1$) in the
thermodynamic limit $N\to\infty$ is most clearly seen by transforming variables
from $(\mathcal{Q}, \mathcal{R})$, Eq.~\eqref{eq:eom}, to
$(\mathcal{Q}, \mathcal{S})$ with
\begin{equation}
  \label{eq:S}
  \mathcal{S} \equiv \frac{\mathcal{R}}{\sqrt{\mathcal{Q}}}.
\end{equation}
The variables $(\mathcal{Q}, \mathcal{S})$ obey another closed set of equations
of motion. After averaging over the inputs using $p(x,y)$ as described in
Section~\ref{sec:dynamics-sde}, we find
\begin{subequations}
  \label{eq:eom-Q-and-S}
  \begin{align}
    \dot{\mathcal{Q}}(t) =& 2+\mathcal{Q}(t) \left(-2k + (2 \lrn_0 -\lrn_0 ^2)\frac{ \mathcal{S}(t) \sqrt{1-\mathcal{S}(t)^2}-\arccos\mathcal{S}(t)}{\pi }\right),\\
    \dot{\mathcal{S}}(t) =& -\frac{\mathcal{S}(t)}{\mathcal{\mathcal{Q}}(t)}+\frac{\lrn_0 ^2
                            \mathcal{S}(t)^2 \sqrt{1-\mathcal{S}(t)^2}}{2 \pi }  +\frac{\lrn_0  \left(1-\mathcal{S}(t)^2\right)^{3/2}}{\pi
                            }-\frac{\lrn_0 ^2 \mathcal{S}(t) \arccos\mathcal{S}(t)}{2 \pi }.\label{eq:eom-S}
  \end{align}
\end{subequations}
Three stream plots of this system for $\lrn_0=2,3,4$, shown in
Fig.~\ref{fig:adatron-dynamics}, reveal a qualitative change in behaviour of the
system ($\mathcal{Q}, \mathcal{S}$) away from a solution with $\mathcal{S}\to1$
and hence $\epsilon\to0$. Indeed, as $\lrn$ increases, $\mathcal{S}\to0$ and
thus $\epsilon\to1/2$. This observation calls for a more detailed analysis of
the system~\eqref{eq:eom-Q-and-S}. Unfortunately, the fixed points of the system
cannot be found explicitly. However, we can consider the limit of small $k$
where the transition is most pronounced, see
Fig.~\ref{fig:error-thermo-vs-nu}. Expanding the equation for
$\dot{\mathcal{Q}}(t)$ around $k=0$ and $\mathcal{S}=1$ shows that
$\mathcal{Q}\sim1/k$ in the steady state. This suggests neglecting the first
term in Eq.~\eqref{eq:eom-S}, which has the appealing consequence of yielding a
closed, single equation for $\mathcal{S}(t)$. This equation has a fixed point
$\mathcal{S}=1$, which is easily checked by substitution. Expanding
Eq.~\eqref{eq:eom-S} around $\mathcal{S}=1$ yields
\begin{equation}
  \dot{S}(t) = \frac{2 \sqrt{2} }{\pi }\left(\frac{\lrn_0 ^2}{3}-\lrn_0
  \right)(1-\mathcal{S})^{3/2} + \mathcal{O}(1-\mathcal{S})^{5/2}
\end{equation}
from which we see that the derivative will change sign at the critical learning
rate $\lrn_c=3$, which is the same value where the well-known breakdown of
AdaTron learning occurs for a setup with $k=0$ and no thermal
noise~\cite{engel2001}. A detailed graphical analysis (not shown) reveals that
the solution $\mathcal{S}=1$ looses its stability at $\lrn_c=3$ while a second
fixed point emerges, which is stable, leading to the collapse of the
generalisation error observed in Fig.~\ref{fig:error-thermo-vs-nu}. The
bifurcation diagram for $\mathcal{S}$ is shown in the right-most plot of
Fig.~\ref{fig:adatron-dynamics}.

\section{Computing the excess heat}
\label{sec:excess-heat}

The most straightforward way to compute the excess heat for learning in the
thermodynamic limit $N\to\infty$ after a quench of the learning rate to
$\lrn(t)=\lrn_0$ is by relying on the machinery developed in
Section~\ref{sec:dynamics-sde}, namely the Itô stochastic differential equation
for the weights, Eq.~\eqref{eq:ito}, which we rewrite slightly here as
\begin{equation}
  \dd \vw = F(\vT, \vw, \vxi) \dd t + \dd \mathbf{W}(t)
\end{equation}
with the total force on the weights $F(\vT, \vw, \vxi)$. The key insight here is
due to K.~Sekimoto, who realised that this
equation is indeed a statement of the first law, with the stochastic heat
increment defined as~\cite{sekimoto1997,sekimoto2015}
\begin{equation}
  \label{eq:heat-definition}
  \ddbar q \equiv  F(\vT, \vw, \vxi) \circ \dd \vw 
  = \frac{1}{2}\left( F(\vT, \vw, \vxi) + F(\vT, \vw+\dd \vw, \vxi) \right)\dd \vw
\end{equation}
where we have evaluated the stochastic product $\circ$ using the Stratonovich or
mid-point convention for every component~\cite{gardiner2009}. For the excess
heat, we replace the total force on the weights with the gradient of the
``non-equilibrium'' potential~\cite{seifert2012}
\begin{equation}
  \label{eq:phi}
  \phi(\vT, \vw; \lrn_0) \equiv - \ln p^\st(\vT, \vw; \lrn_0)
\end{equation}
where $p^\st(\vT, \vw; \lrn_0)$ is the steady-state distribution for $\lrn_0$ with
steady-state values $\mathcal{Q}^\st(\lrn_0)$ and $\mathcal{R}^\st(\lrn_0)$. Hence
after averaging over the thermal noise, we find for the average increment in
excess heat of the system over $N$
\begin{align}
  \ddbar Q^\ex_n = \frac{1}{N} \ddbar Q^\ex \equiv& -\frac{1}{N}\dd \vw \circ
                                                    \nabla_\vw \phi(\vT, \vw; \lrn_0)  \\[1em]
  =&-\frac{1}{N}\dd \vw \circ
     \frac{\vw-\mathcal{R}^\st(\lrn_0)\vT}{\mathcal{Q}^\st(\lrn_0)-\mathcal{R}^\st(\lrn_0)^2}\\[1em]
  \begin{split}
    =& \frac{1}{\mathcal{Q}^\st(\lrn_0)-\mathcal{R}^\st(\lrn_0)^2}[ k
    (\mathcal{Q}(t)-\mathcal{R}^\st(\lrn_0)\mathcal{R}(t)) -\lrn_0
    \avg{\sgn(x)y\mathcal{F}(x,y)} \\[.8em]
    & \qquad + \lrn_0 \mathcal{R}^\st(\lrn_0)\avg{\sgn(x)x\mathcal{F}(x,y)}
    -\lrn_0^2\avg{\mathcal{F}(x,y)^2}/2-1 ]\dd t.
  \end{split}
\end{align}
where $\avg{\cdot}$ now indicates an average over the distribution $p(x, y)$ as
described in Appendix~\ref{sec:dynamics-sde} and we remind ourselves that we
chose units where $D=1$.
\end{widetext}

\bibliography{Rules}

\end{document}